\documentclass[12pt]{JHEP}
\usepackage{amsmath,epsfig}
\usepackage{amssymb,amsfonts}
\usepackage{latexsym}
\usepackage{epsfig}

\relax
\renewcommand{\theequation}{\arabic{section}.\arabic{equation}}
\def\be{\begin{equation}}
\def\ee{\end{equation}}
\def\bea{\begin{eqnarray}}
\def\eea{\end{eqnarray}}

\newcommand\fverb{\setbox\pippobox=\hbox\bgroup\verb}
\newcommand\fverbdo{\egroup\medskip\noindent%
                        \fbox{\unhbox\pippobox}\ }
\newcommand\fverbit{\egroup\item[\fbox{\unhbox\pippobox}]}

\newcommand{\bear}{\begin{eqnarray}}

\newcommand{\eear}{\end{eqnarray}}
\def\hri#1#2{\href{http://arxiv.org/abs/#1}{[ArXiv:#1][#2]}}
\def\hre#1#2{\href{http://arxiv.org/abs/#1/#2}{[ArXiv:#1/#2]}}
\def\hspi#1#2{\href{http://www.slac.stanford.edu/spires/find/hep/www?irn=#1}{#2}}
\newbox\pippobox
\def\dc{\delta c}
\def\ls{\ell_s}
\def\ie{{\it i.e.~}}
\def\lab{\label}
\def\6{\partial}
\def\f{\Phi}
\def\a{\alpha}

\def\half{\frac12}

\def\le{\left}
\def\ri{\right}
\def\cO{{\cal O}}

\def\pa{\partial}

\def\co{{\cal O}}
\def\e{\epsilon}
\def\m{\mu}
\def\n{\nu}

\def\s{\sigma}

\def\sp{\;\;\;,\;\;\;}

\def\z{\zeta}
\def\sq
\def\a{\alpha}
\def\b{\beta}
\def\l{\lambda}

\def\tr{{\rm Tr}}
\def\k{\chi}
\def\tb{\overline{t}}
\def\L{\Lambda}
\def\bb{\overline{b}}
\title{Exploring improved holographic theories for QCD: Part I}

\author{U. G{\"u}rsoy$^{1,2}$,
\href{http://hep.physics.uoc.gr/~kiritsis/}{E. Kiritsis}$^{1,3}$\\
$^1$\href{http://cpht.polytechnique.fr/cpht/cordes/}{CPHT, Ecole Polytechnique, CNRS,
 91128, Palaiseau, France}\\
 ( UMR du CNRS 7644).\\
~\\
$^2$\href{http://www.lpt.ens.fr/}{Laboratoire de Physique Th\'eorique},\\
Ecole Normale Sup\'erieure,
24, Rue Lhomond, Paris 75005, France.\\
~\\
$^3$\href{http://hep.physics.uoc.gr/}{Department of Physics, University of Crete,
71003 Heraklion, Greece}\\
~\\}

\preprint{ArXiv:0707.1324 \\ CPHT-RR027.0507}      

\abstract{Various holographic approaches to QCD in five dimensions are explored using input
both from the putative five-dimensional non-critical string theory as well as QCD. It is argued that a gravity theory in five dimensions coupled
to a dilaton and an axion may capture the important qualitative features of pure YM theory. A part of the effects of higher $\a'$-corrections
is resummed into a dilaton
potential. The potential is shown to be in one-to-one correspondence with the exact $\beta$-function of QCD, and its knowledge determines
 the full structure of the vacuum
solution. The geometry near the UV boundary is that of AdS$_5$ with logarithmic corrections reflecting the asymptotic freedom of QCD.
We find that all relevant confining backgrounds have an IR singularity of the ``good" kind that allows unambiguous  spectrum computations.
Near the singularity the 't Hooft coupling is driven to infinity. Asymptotically linear glueball masses can also be achieved.
The classification of all confining asymptotics, the associated glueball spectra and meson
dynamics are addressed in a companion paper \href{http://arxiv.org/abs/0707.1349}{ArXiv:0707.1349} }


\begin{document}

\def\g{\gamma}
\def\go{\g_{00}}
\def\gi{\g_{ii}}

\maketitle 

\section{Introduction and conclusions}

The correspondence between large-$N_c$ gauge theory and string theory first advocated by 't Hooft \cite{hooft} has taken a novel
(and mostly unexpected) form
after the Maldacena conjecture, \cite{malda}.
Although the most precise version of the correspondence was formulated for the (conformally invariant)
maximally supersymmetric Yang Mills theory in four dimensions,
in several other cases, holographic techniques proved useful, and several gravitational duals describing confining theories in the IR
have been proposed.
One concrete application of holographic duality is to establish a novel quantitative understanding of QCD, in particular of phenomena
where strong IR physics is important. These include confinement, chiral symmetry breaking, as well as quantitative issues about the meson
and baryon spectrum and their interactions.

Critical holographic models obtained as solutions of ten-dimensional string theory \cite{D4,MN,KS}
have been successful in their qualitative description
of confinement and other IR dynamics, including a semi-quantitative agreement of the
glueball spectrum \cite{glue} and thermal properties \cite{D4}.
At the same time, the theories contain Kaluza-Klein  modes, not expected in QCD,
with KK masses of the same order as the dynamical scale
of the gauge theory. Above this scale the theories deviate from QCD. In the solution associated to non-supersymmetric wrapped $D_4$ branes, \cite{D4}
the UV completion should be thought of as a 6D theory on  M5 branes.
In the Chamsedinne-Volkov solution\cite{cvo}, interpreted holographically by Maldacena and Nunez, \cite{MN}, the theory becomes the
6d theory obtained by wrapping
NS$_5$ branes on a two-sphere. Finally in the Klebanov-Strassler solution, \cite{KS} the theory passes through a
large set of Seiberg dualities to end up with a non-conformal quiver theory in the UV.

An obvious way to avoid the extra KK modes is to consider non-critical string theories.
Indeed, if our general intuition
about  holography is correct, QCD should be described by a string theory living in 5 dimensions.
However, it is expected from general arguments that non-critical string duals of large-$N_c$
gauge theories will have background curvatures
and other invariants of the same order as the string scale and an $\alpha'$ expansion is
not expected to be a priori a reliable approximation.
Such an expectation is partly based on the fact that the part of the potential due to the
central charge deficit is of order ${\cal O}(1)$ in $\ell_s^{-2}$ units.
One may also give a more rigorous statement \cite{BCCKP} for supersymmetric theories,
based on the conformal anomaly \cite{hs}.

Despite the hostile environment  of  non-critical theory, several attempts have been made to
understand holographic physics in this regime,
based on two-derivative gravitational actions, \cite{ks,km,BCCKP,cr}. Moreover, such attempts were seconded
by studies of the physics of probe branes in
non-critical
backgrounds \cite{f,fnp,amt} and non-critical orientifolds \cite{in} that provide
 side information on the holographic approaches.
Most of these attempts focused on 4d CFTs with  ${\cal N}=1$ supersymmetry \cite{km,BCCKP}
or without it \cite{ks,BCCKP,cr}.
The rational has been, that although $\alpha'$ corrections are generically expected to  be
substantial, qualitative information should be (mostly) reliable.
Moreover, the high symmetry may guarantee that some quantities could be reliably computed.
Part of this intuition stems from exact solutions to
2d WZW CFTs where ratios of conformal weights can be calculated classically, and are exact to
all orders in the $\alpha'$ expansion.

A different and more phenomenological approach was in the meantime developed, and is now known as AdS/QCD.
The original idea was formulated in \cite{ps} and it was successfully applied to the meson
sector in \cite{adsqcd1,adsqcd2}.
The bulk gravitational background consists of a slice of $AdS_5$, and a constant dilaton.
There is a UV and an IR cutoff. Moreover, the confining IR physics is imposed by boundary
conditions at the IR boundary.
This approach seems very crude and indeed it is,
when applied to  the pure glue sector. However it has
been partly successful in studying meson
physics despite the fact that the dynamics driving chiral symmetry breaking must be
 imposed by hand via IR boundary conditions.

In this paper and its companion \cite{part2}, we will investigate a mostly phenomenological
approach that runs somewhere in-between
non-critical string theory approaches and AdS/QCD. On the one hand we will investigate and
 motivate what kind of effective theory we expect to
describe QCD, based on our understanding of string theory.
On the other hand we would like to match this  with what  we expect from QCD in the UV and the IR.

There is intriguing evidence from  studies of QCD, that high-dimension operators, that should
 be associated to stringy modes, are not very important
for (some of the) physics at short distances. This is suggested by the success of SVZ sum rules, \cite{SVZ}
 which tie together the UV physics with the IR physics in QCD.
A counter-argument relies on the fact that as we understand from critical holography,
the large 't Hooft coupling limit $\l\gg 1$ is necessary to suppress
higher $\alpha'$ corrections. Therefore, in QCD, as this coupling is driven to zero in the UV,
such corrections will become dominant.
 However, as  we argue below, progress can be achieved in this direction despite the difficulties.

Pure 4d YM  at large $N_c$ is expected to be dual to a string theory living in 5 dimensions. The relevant
low-lying fields are expected to be dual to the lowest
dimension operators, namely the graviton (dual to  $T_{\m\n}\sim Tr[F_{\m\n}^2-{1\over 4}\delta_{\m\n}F^2]$), the dilaton $\phi$
(dual to $Tr[F^2]$)
 and the RR axion $a$ (dual to $Tr[F \wedge F])$.
Moreover, in the string frame, the theory has a very simple dilaton potential $\sim \delta c$
that reflects the fact that
the associated string theory is non-critical.
The theory however must contain a RR four-form whose flux  seeds the $D_3$ branes and
therefore generates the $U(N_c)$ gauge group.
The presence of such a form is compatible with the spectrum of operators we advocate, as
in 5 dimensions a four-form is non-propagating.
By integrating out the four-form we generate new terms in the dilaton potential.
This is shown in section \ref{motv} and appendix \ref{dilpot}.
Therefore, taking into account higher $\alpha'$ terms involving the four-form only,
generates a non-trivial potential for the
dilaton $\phi$. Hence, the special nature of the four-form implies that its field strength, although
it carries powers of $\alpha'$, effectively carries no derivatives, and it should therefore be incorporated in the potential.

There are two properties that this potential satisfies. The first is that it is a nontrivial
 function of the 't Hooft coupling $\lambda\sim N_c e^{\phi}$, ($\l$ thus defined,   is expected
to remain finite in the large-$N_c$ limit).
The second is that (a) it has a regular expansion around $\lambda=0$ (b) it vanishes as
 $\lambda^{4\over 3}$ at $\lambda=0$ (in the Einstein frame).
To next to leading order we expect
\be
V(\lambda)=\lambda^{4\over 3}\left(\delta c-\l^2+\cdots\right)
\label{i1}\ee
This potential was established in \cite{BCCKP} and is generic in dimensions 5, 4 and 3.
The fact that it vanishes at $\l=0$ implies that the asymptotic geometry in the UV
does not approach AdS$_5$, as we might expect. This is analyzed in detail in appendix \ref{exponential}.
However the potential above has another intriguing property: it has an AdS minimum at a finite value of the
't Hooft coupling, and therefore there is a related AdS$_5$ solution with fixed finite  't Hooft coupling.
The expectation that this AdS$_5$ may be the UV limit of a non-trivial solution with (logarithmically)
 running $\l$ seems however to fail as explained in appendix \ref{pertpot}.
Therefore the only ``correct" weak-coupling asymptotics of the dilaton potential must satisfy
\be
\lim_{\lambda\to 0} V(\lambda)=V_0\not=0
\label{i2}\ee
This guarantees the existence of a solution which is asymptotically AdS$_5$ near the UV
 boundary, provided the 't Hooft coupling vanishes there.
Moreover, if we wish that in the same UV region, the 't Hooft coupling runs logarithmically
with energy, then the weak coupling expansion of the potential
must be of the form
\be
V(\lambda)=V_0\left(1+\sum_{n=1}^{\infty}{V_n}(\lambda^{a})^{n}\right)
\label{i3}\ee
with $a$ some positive real number that can be shifted to the wave function renormalization of the dilaton.
As suggested by the perturbative QCD $\beta$-function we will select $a=1$.
As we show in section \ref{motv} and appendix \ref{afree}, this expansion of the potential
is equivalent to the perturbative expansion of the QCD $\beta$-function and
the coefficients $V_n$ can be directly related to the
perturbative $\beta$-function coefficients.

There are two obvious questions that accompany the discussion above:
\begin{enumerate}
\item What is the origin of the non-zero constant $V_0$?

\item In the arguments above we have neglected higher $\a'$-corrections
involving the curvature and derivatives of the scalars.

\end{enumerate}

As we indicate in appendix \ref{hd}, a non-zero constant $V_0$ in the effective dilaton
potential may be generated from the higher-curvature corrections. Although this also implies
 that  higher $\alpha'$-corrections cannot be neglected after all, we would like to take the
bold step and assume that the only quantitatively important part of this class of $\alpha'$
 corrections is just to generate a non-zero $V_0$ as well as the rest of the terms of the dilaton potential.
This step, has as a consequence that we will eventually deal with a two-derivative
 effective action but with a general dilaton potential satisfying (\ref{i3}) at weak 't Hooft coupling.
Although this approach cannot be rigorously defended, our attitude is exploratory.
An extra motivation
comes from the success of the SVZ sum rules \cite{SVZ}.
Moreover, as we
 show below, our results are encouraging.

To summarize the discussion above, our starting point is an action (in the
 Einstein frame) of the form
 \be
   S=M^3N_c^2\int d^5x\sqrt{g}\left[R-{4\over 3}{(\partial\l)^2\over \l^2}-
{Z(\l)\over 2N_c^2}(\partial a)^2+V(\lambda)
\right]
    \label{i6}
\ee
with the potential $V(\l)$ having around $\l=0$ an asymptotic expansion of the form
(\ref{i3}).

The axion kinetic term in (\ref{i6}) deserves  some comments. In a way similar with the potential,
 we expect a non-trivial function of the dilaton
$Z(\l)$ multiplying its kinetic term.\footnote{A priori, similar functions might
also multiply the Einstein term and the dilaton kinetic term.
Such terms can however be removed by Weyl-rescaling of the metric and redefining $\l$.}
Moreover, as explained in detail in appendix \ref{axion-corr}, the axion kinetic term
is of order ${\cal O}(1/N_c^2)$ compared with the rest of the terms
in (\ref{i6}). This is due in string theory to the fact that the axion is a RR field
and therefore has a suppressed dilaton dependence.\footnote{This is not valid for the RR
four-form as its field strength is linear in the number of colors $N_c$.}
The same argument indicates that terms involving higher powers of $(\partial a)^2$
will be further suppressed at large $N_c$.
Therefore, the $\a'$-expansion of the axion terms is effectively an  $1/N_c$ expansion.
The perturbative asymptotics of $Z(\lambda)$ therefore should be
\be
Z(\lambda)=Z_a+{\cal O}(\lambda^2)\sp \l\to 0
\label{i7}\ee
String theory gives a $\l^2$ contribution to leading order. However as in
the case of the potential, higher derivative corrections are expected
to generate also a constant piece. This turns out to be in agreement with
 perturbative QCD.

As shown in section 4 of \cite{part2}, $Z(\lambda)$ determines an analogue
 of a  $\beta$-function for the QCD $\theta$-parameter.
This interpretation has however caveats that are discussed in the same section.
It is obvious that while searching for the solution of the equations of
motion stemming from
the action (\ref{i6}), that will describe the QCD vacuum, the axion contribution
 can be neglected
to leading order in $1/N_c$. Once this solution is found, the axion equation
of motion can be solved in order to determine the
profile of the QCD $\theta$-parameter and its associated physics.

 Therefore, to leading order in $1/N_c$, the vacuum structure of QCD, is
 captured by  a solution of (\ref{i6}) with the appropriate
$AdS_5$ asymptotics. All the properties of the 4D gauge theory
depend on a single function of the 't Hooft coupling,
the superpotential $W(\l)$,  defined as\footnote{Generically, passing from
 the potential to the superpotential and the associated
first order equations involves an arbitrary constant that can be thought
of as the single constant of integration of (\ref{i8}).
However in our case it turns out that the relevant solution is special,
and does not have this extra dependence on an arbitrary constant.
This is explained in appendix E of \cite{part2}.} :
\be
V(\l) =
-{4\over 3}\l^2\left({ d W\over d\l}\right)^2 + {64\over 27} W^2.
\label{i8}\ee
 This dependence can be mapped (up to the overall
AdS length $\ell$, related to $V_0$  by $V_0={12\over \ell^2}$) to the
(exact) $\beta$-function of QCD, $\beta(\l)$.
In particular, for small $\l$, all coefficients of the superpotential can be mapped to the perturbative
terms of the $\beta$-function.
Concretely, the following relation holds:
\be
\beta(\l)=-{9\over 4}\l^2~{d\log W(\l)\over d\l}
\label{i9}
\ee
Moreover fitting to QCD data we learn that, \cite{part2}
\be
\ell\simeq 6.26 ~\ell_s
\ee

We expect three integration constants for the equations of motion of the metric and $\l$.
One of them fixes $W(\l)$ as a solution of eq. (\ref{i8}). We show in
\cite{part2} that this integration constant is completely fixed by using asymptotic freedom as an
input from the gauge theory. Among the second and third integration constants that arise from
the first-order differential equations for $\l$ and $A$, only a single combination remains due to the
reparametrization invariance. This remaining
integration constant amounts to a definition
of $\Lambda_{QCD}$.
>From eq. (\ref{i9}) it is apparent that
the exact $\beta$-function determines completely the geometry up to a choice
of $\Lambda_{QCD}$.

One  result, in our two-derivative approach, concerns the investigation
of potential confining backgrounds in the IR.
We choose the conformal coordinate system and write the Einstein metric as
\be
ds^2=e^{2A(r)}(dr^2+\eta_{\m\n}dx^{\m}dx^{\m})
\label{i10}
\ee
with the AdS$_5$ boundary at $r=0$.
As we show,  $e^{A(r)}$ monotonically decreases from $\infty$ at the UV boundary,
to $0$, or to a finite value, in the IR (see section 3.2). We therefore
take the scale factor in the Einstein frame as our definition
for the energy of the gauge theory.
\be
E=e^{A(r)}
\label{i11}
\ee
 In the case when the Einstein frame scale factor remains finite at the
IR singularity, this indicates that the dual theory is defined only above
a certain  energy. Such models however fail to satisfy some of the
properties that are believed to  hold in confining theories,
namely the screening of the magnetic color charges.

 In our analysis of the non-perturbative
regime, we give a general classification of possible IR geometries
according to their confining properties. We use, as a characterization
of confinement, the Wilson loop area law:
by ``confining,'' we label those backgrounds for which the QCD
string (identified with the fundamental string) has a finite
tension. We then analyze various properties of confining backgrounds.
Here is a summary of our findings:

\begin{itemize}

\item We provide a relation between the  $\beta$-function of the gauge theory defined
by an infinite series expansion in the 't Hooft coupling, and the scalar potential (or rather the superpotential)
of the gravitation theory defined by a similar expansion in the dilaton.

\item We study certain $\alpha'$ corrections to the scalar potential. In particular, we find that
in the identification of the $\beta$-function and scalar potential coefficients, the first two
scheme-independent coefficients of the $\beta$-function receive no $\alpha'$ corrections.
Moreover we find that the $\alpha'$ corrections to the identification of the energy scale with the scale factor of the metric
can be set to zero in a particular scheme for computing the higher order $\beta$-function coefficients, $b_2,b_3,\dots$.

\item We show that {\em all} confining backgrounds have a
singularity in the Einstein metric at some value $r_0$ of the $r$ coordinate.
There are two distinct cases for the position $r_0$ of the IR singularity.
One possibility is finite $r_0$. The other is $r_0=\infty$.
The IR singularity is always of the ``good kind" \cite{gubser}: fluctuation spectra of
 various fields are well defined and are not affected directly by the presence
of the singularity.

\item For regular dilaton potentials\footnote{We assume that the
potential, hence the $\beta$-function, do not have singularities
at finite $\lambda$.}, the 't Hooft coupling $\l$ always becomes
infinite at the IR singularity.

\item  In the case $r_0=\infty$, the string frame metric is not only regular at the IR singularity but its curvature also  vanishes.
Put otherwise, in the string frame,
 the IR singularity is only due to the diverging 't Hooft coupling constant.
{ This suggests that the supergravity approximation may be a good approximation in the
IR region.}

\item We classify all superpotentials $W(\l)$ that give rise to confining backgrounds:
We parametrize their asymptotics for $\l\to \infty$   as\footnote{Since
our results are continuous in  the parameters $P$ and $Q$,
our classification also extends to any functions $W(\l)$ that
has a well-defined limit for $\l \to \infty$.}:
\be
W(\l)\sim (\log\l)^{P\over 2}~\l^Q\sp P,Q\in R.
\label{i12}
\ee
The 't Hooft couplings diverges in the IR as
\be
\l\sim E^{-{9\over 4}Q}\left(\log{1\over E}\right)^{P\over 2Q}, \qquad E\to 0.
\label{i17}\ee

\begin{enumerate}

\item
${ Q> 2/3}$ or  ${ Q = 2/3}$ and  ${ P > 1}$
leads to confinement and a singularity at finite $r=r_0$. The
scale factor  $e^A$ vanishes there as
\be
 e^A(r) \sim \left\{\begin{array}{ll} (r_0-r)^{4\over 9Q^2 -4} &\quad Q>{2\over 3}
  \\ \exp\left[-{C\over (r_0 -r)^{1/(P-1)}}\right] & \quad Q = {2\over 3} \end{array}
\right. \ ,
 \label{i12-2}
\ee
where $C$ is a positive constant related to the integration constants.
\item
${Q = 2/3}$, and  ${ 0 \leq P < 1}$ leads to confinement  and a singularity at $r=\infty$
The scale factor $e^A$ vanishes there as
\be\label{i12-3}
e^A(r) \sim \exp[-Cr^{1/(1-P)}].
\ee

\item  ${ Q = 2/3, P = 1}$ leads to confinement but the singularity may be at a finite or
 infinite value of $r$ depending on subleading asymptotics of the superpotential.

\end{enumerate}

 The above exhaust  all cases that confine.
All other cases ($Q<2/3$, or $Q=2/3$ and $P<0$) fail the Wilson loop test.
 By eq. (\ref{i9}),
this classification directly relates the confining property to the IR
behavior of the QCD $\beta$-function.

\item  If $Q< 2\sqrt{2}/3$,  no {\em ad hoc}
 boundary conditions are needed but UV and  IR normalizability
completely determines
the glueball spectrum. This is unlike standard AdS/QCD and other approaches.
 Since the spectrum is completely determined  from the geometry,
and, as discussed above, the latter is in one-to-one correspondence with the $\beta$-function,
our construction provides a direct link between the $\beta$-function and the spectrum. On the other hand,
 when $Q> 2\sqrt{2}/3$, the spectrum is not well defined without extra boundary conditions in the IR because
both solutions to the mass eigenvalue equation are IR normalizable. \footnote{In \cite{cr} an
 exact solution was studied based on the potential $W(\l)=W_0+W_1 \l^{4\over 3}$
in our normalizations. This generates logarithmic running in the UV and corresponds
to a confining background in the
IR. However, as it corresponds to $Q=4/3$, extra boundary conditions are needed at the
IR singularity. The spectrum heavily depends on the choice of these boundary  conditions.}

\item For all potentials  that confine,  the spectrum of $0^{++}$ and $2^{++}$ glueballs  has a mass gap.
Moreover, except for the borderline case $Q=2/3$, $P=0$\footnote{
This  is also exactly the linear dilaton background in the IR.
The spectrum of glueballs is however continuous in this case. Because of this, this case is
considered no further.}, the spectrum is also purely discrete.
For the $0^{+-}$ glueballs an extra assumption is needed about the strong-coupling
 asymptotics of the function $Z(\lambda)$ in (\ref{i6}):
\be
Z(\l)\sim \l^d\sp d>2~~~{\rm as}~~~\l\to \infty.
\label{i13}\ee
If this is satisfied, the  $0^{-+}$ spectrum is also gapped and discrete.
We find that in QCD $d=4$.

\item In all the physically interesting confining backgrounds (i.e. those
that do not require extra boundary conditions in the IR), the magnetic
color charges are screened. This is shown by studying the D$_1$ branes embedded in the geometry, see \cite{part2}.
This is an improvement with respect to AdS/QCD models,
where magnetic quarks are also confined instead of being screened.

\item Of all the possible confining asymptotics, there is a unique one that guarantees
``linear confinement" for all glueballs.
It corresponds to the case $Q=2/3, P=1/2$, i.e. strong-coupling
superpotential and $\beta$-function asymptotics:
\be
W(\l)\sim (\log \l)^{1\over 4}~\l^{2\over 3}\sp \beta(\l)=-{3\over 2}\l\left[1+{3\over 8\log \l}+\cdots\right]
\label{i14}\ee
In this case the 't Hooft coupling diverges  with energy as
\be
\lambda\sim  E^{-{3\over 2}}\left(\log{1\over E}\right)^{3\over 8}
\label{i15}\ee
 in the IR.
This choice also seems to be preferred from considerations of the meson sector as discussed below.

\item The meson sector, assuming $N_f\ll N_c$, is implemented by using $N_f$ pairs of $D_4-\bar D_4$ branes
 embedded in the 5-dimensional
background along the lines first pointed-out in ten dimensions in  \cite{ss}. Moreover, the open string tachyon field is
included and it is dual to the scalar and pseudoscalar
quark bilinears as first advocated in \cite{ckp}.
In section 5 of \cite{part2} we have studied the non-linear equation that determines the tachyon profile (vev).
It was shown in \cite{ckp} that consistency with the anomaly structure implies that the tachyon
field must diverge before or at the IR end of space.
What we find here is that for the confining background the tachyon necessarily diverges
at the IR singularity, signaling chiral symmetry breaking
and the IR recombination of the flavor branes.
Moreover, it is found that although the tachyon cannot diverge before the IR singularity,
its derivatives do generically diverge. Such a divergence
is physically unacceptable. Imposing its absence, determines the quark vacuum condensate
in terms of the UV quark masses that act as sources for the
tachyon field.

\item As in \cite{ckp}, the spectrum of mesons exhibits (almost) linear confinement due to the tachyon potential
rather that the graviton-dilaton background.
We have also studied the masses of the simplest mesons, the vectorial ones which are
independent of the details of chiral symmetry breaking.
They depend in general on a combination of the
confining QCD scale (appearing in the graviton-dilaton data) and the
 AdS length $\ell$. We find that the special background advocated above, corresponding to $P=1/2$,
which provides linear confinement in the glueball sector
is also the one in which meson masses do not depend on the AdS$_5$ scale,
 as expected on general grounds.
Therefore we obtain a simple linear relation between the mass scales of
 glueballs and generic mesons (this does not include the
pseudo-Goldstone bosons).

\item We calculate numerically the $0^{++}$, $0^{-+}$  and $2^{++}$ glueball
spectra for
 $\beta$-functions that interpolate between
 the standard perturbative QCD regime and
the confining regime, both for the $r_0$ finite and infinite cases.
We compare the glueball spectra with lattice results.
Although there is no universal consensus
on the reliability of various lattice results and their relationship to the
large-$N_c$ limit, we attempt a comparison with various confining IR asymptotics.
We find that the cases of $r_0=\infty$ are preferred by the data.

\item We analyze the axion sector of the theory in section 4 of \cite{part2}. We solve
the equation for the axion to find
\be
a(r)=({\theta_{UV}}+2\pi k){\int_r^{r_0} {dr \over e^{3A}Z(\l)}
\over \int_0^{r_0} {dr \over e^{3A}Z(\l)}}
\label{axiona}\ee
where $\theta_{UV}$ is the UV value for the QCD $\theta$-angle and the integer $k$
labels different large-$N_c$ vacua.
In particular we reproduce the well known result that the $\theta$-dependent
 vacuum energy is to leading order in $1/N_c$ proportional to $Min_k(\theta_{UV}+2\pi k)^2$
and relate the coefficient (topological susceptibility) to the standard QCD $\beta$-function
as well as the axion $\beta$-function $Z(\l)$.
One important corollary of our analysis is that $a(r)$,
``the effective $\theta$-angle" vanishes as a power of the energy
in the IR,
\be
\theta_{eff}(E)\sim E^3(\log E)^{1\over 2}
\ee
 This is an interesting result as it indicates that the YM dynamics screens the UV $\theta$-angle.

\end{itemize}

Before continuing we will comment on what we hope to achieve with this approach.
As it will become evident, this phenomenological approach does not have at this point the status of
well controlled approximation. Therefore at best we can hope to achieve the following:

\begin{enumerate}

\item To provide a successful and predictive phenomenological model
with few parameters. A good analog of such a model is the
Lund Monte Carlo that describes hadronization in high energy collisions
 using a basic string model and a  dozen parameters.

\item If item 1 turns out to be successful, such a model can provide intuition and data, to guide
serious searches for constructing a string theory for QCD.

\item It may provide hints for new phenomena in the theory. A good
such example may turn out to be our observation that
the IR $\theta$-angle in large-$N_c$ QCD vanishes.

\end{enumerate}

This introduction summarizes the present paper as well its companion
\cite{part2} that should be read as a natural continuation of this
one. The structure of this paper is as follows.

In the next section
we describe in detail the gravitational set up used to explore the
duals of QCD-like gauge theories. In particular we analyze the
general form of the scalar potential.

In section \ref{gfgt}, we analyze the
equations of motion for the coupled scalar-gravity system and derive
the precise relation between the full $\beta$-function of the gauge
theory and the scalar potential.

In section 4, we derive the UV
asymptotics of the solutions close to  the $AdS$  boundary,
using the gauge theory input, namely asymptotic freedom.
 Then,  we discuss  the qualitative features of a
class of  IR asymptotics in the deep
interior of the geometry that lead to confinement.
 This section also
includes a detailed discussion of the $\alpha'$ corrections near the
boundary. In particular, we show that in the expansion near the UV,
the $\alpha'$ corrections are always accompanied with the scheme-dependent
$\beta$-function coefficients.

Section 5 is devoted to
specific examples of geometries that follow from a certain choice of
the scalar potential. In section 5.1 we present a simple geometry
that displays asymptotic freedom in the UV and linear confinement in
the IR. This example is used in \cite{part2} to compute the glueball
spectra. In section 5.2 we present a sample geometry that displays a
Banks-Zaks type fixed point in the IR.

Section 6 investigates the
fluctuations of the various fields in the geometry and how the dilaton potential
(ie. the perturbative $\beta$-function) modifies the fluctuation equations near
the UV boundary.

Various appendices detail our computations. In appendix \ref{afree}, we
present a detailed analysis of the expansion near the UV. In
appendix \ref{dilpot} we analyze a general form for the action of a
dilaton-axion-gravity system that incorporates the $\alpha'$
corrections by assuming general forms in the kinetic terms for these
fields. We derive general solutions of the action and determine the
$\alpha'$ corrections (in the UV), to the various quantities used in
the bulk-boundary identification. Appendix \ref{pertpot}  describes the geometry
that follows from the naive effective potential which ignores the
$\alpha'$ corrections and studies the fluctuations around the AdS
vacuum of this potential. In appendix \ref{exponential}, we analyze the solutions to
the coupled Einstein-scalar system with an exponential potential. We
derive and classify all the solutions and describe the fixed points.
Finally, appendix \ref{bapp} presents another example of a geometry that
approaches to a conformal fixed point in the IR with an exponential
tail in the $\beta$-function.

As Part II of this work is an essential sequel of this paper we briefly review here its structure.
It is mainly   devoted to the analysis  of the non-perturbative
regime of our construction, i.e. to the IR properties of
the 5D geometry. There,   the general classification
of IR asymptotics which lead to confinement is given.
The qualitative features of the glueball spectra are discussed, and
a relation is established between the existence of a mass gap and
 the confining property of the QCD string.
The dependence of
the spectrum on excitation number is discussed,
and the IR asymptotics  that lead to a linear glueball spectrum are  identified.
The  analysis is extended to mesons, indicating that the setup
provides a concrete realization of the holographic implementation
of chiral dynamics proposed in \cite{ckp}. The properties
of the 5d axion are discussed, and what they may imply  for the QCD $\theta$-parameter.
Finally, numerical computations of glueball spectra
in concrete models are performed. The models are  defined in terms of an exact   $\beta$-function
that interpolates
between the desired UV and IR asymptotics. The detailed
structure of the Part II  can be found in its introduction.

\section{Motivating the gravitational dual of a large-N$_c$ gauge theory}
\lab{motv}

The gauge theories of interest in this paper, are four-dimensional U(N$_c$) gauge theories at large $N_c$.
In particular, we assume the presence of no further adjoint fields, and therefore, the holographic dual theory is expected
to live in five dimensions. Fundamental matter can be present but we will assume here that the number of flavors $N_f\ll N_c$.
Therefore, quarks can be eventually incorporated as four-brane probes inside the five-dimensional geometry.

\subsection{The spectrum}

The relevant non-critical string is therefore five-dimensional\footnote{For
a similar  discussion of the  spectrum, see \cite{aldo}}.
As there are no fermionic gauge-invariant operators in pure YM theory
we do not expect the associated string theory to have space-time fermions,
(No NSR and RNS sectors).

>From the NSNS sector we will obtain as low lying fields a metric $g_{\m\n}$, a dilaton,
 $\phi$ and a two-index antisymmetric tensor $B_{\m\n}$.
The generic bulk fields should be in one-to-one correspondence
with the low-dimension gauge-invariant operators of the gauge
theory.
In particular
 the (classically) traceless stress tensor
$T_{\m\n}=Tr[F^2_{\m\n}-{1\over 4}\eta_{\m\n}F^2]$ is dual to the
five-dimensional graviton $g_{\m\n}$. $Tr[F^2]$ is dual to the
dilaton $\phi$.
$B_{\m\n}$ is expected to be dual, in analogy with ${\cal N}=4$ sYM, to
a higher dimension operator of the form $Tr[d_{abc}F^aF^bF^c]$ \cite{f3}.
As it will be trivial in the vacuum solution, due to Lorentz invariance, we will neglect it in the sequel.
It is however expected to generate a tower of $1^{-+}$ glueballs in the theory.

The RR sector massless fields are summarized into the tensor product of two 5-dimensional spinors.
The product of two five-dimensional spinors gives the fields strengths $F_0,F_1,F_2,F_3,F_4,F_5$.
There is an automatic  duality condition for this product that relates $F_5\sim F_0$, $F_4\sim F_1$ and $F_3\sim F_2$.
Therefore only $F_0,F_1,F_2$ are independent.

$F_1=dC_0$ generates an axion field $C_0=a$ that is dual to  $Tr[F\wedge F]$ in the YM theory.
 The dual
of the axion field-strength is a four-form field strength
$F_4$ and the associated three-form $C_3$  couples to domain
walls (that separate different $k$-vacua).

$F_5=dC_4\sim F_0$
generates a four-form that seeds the $D_3$ branes responsible for the $U(N_c)$
gauge group. Its dual is a zero-form field strength that couples to
bulk instantons.

Finally $F_2=dC_1$ generates a vector and its presence seems puzzling as there is no candidate YM operator that it is dual to.
It turns out however that such a form couples minimally to the baryon density on flavor branes \cite{ckp},
and its presence is therefore justified.
It is certainly trivial in the QCD vacuum as it will break four-dimensional Lorentz invariance otherwise.
Its kinetic term is suppressed by $1/N_c^2$ as that of the axion.
It can however be used to turn non-trivial baryon number densities in QCD.
It is expected to generate a tower of spin-1 glueballs comparable in profile to the $0^{-+}$ ones.
We will analyze it further in \cite{part2}.

We would like at this point to stress that the presence of RR fields is important with matching with the behavior expected from QCD.
In particular, the special dependence on the dilaton of terms in the effective action containing RR fields (in particular the axion),
 is in complete agreement with the large $N_c$ suppression we expect for such terms in the gauge theory.

Flavor is expected to be generated by space filling $D_4+\bar D_4$ brane pairs filling (most of) the five dimensional bulk.
They couple individually to a $C_5$ form. Overall neutrality is needed however as it equivalent to anomaly
cancellation in the YM theory.\footnote{This appears to be rather general in holographic setups.
Cancellation of bulk RR charges is equivalent to the absence of gauge anomalies of the large-$N_c$ boundary theory.}
As the branes are space-filling, the coupling to $C_5$ will not play a role.
 Subleading couplings to other RR fields are essential though and will be
discussed in \cite{part2}.

The structure of the RR sector is similar to what we would obtain from an orientifold of the 0B theory
 \cite{polya,carlo,in} compactified to
5 dimensions.
Several authors advocated
the relevance of a closed string tachyon in the vacuum structure of
QCD motivated by the ten dimensional type-0 string theory.
 When $N_f=0$ we see no place for such a closed string tachyon,
as there is no such low-dimension gauge-invariant operator in the
gauge theory. If type-0 theory has any connection to pure YM theory, then the ten-dimensional tachyon must become rather
massive in the five-dimensional theory.
Of course if we add $N_f$ flavor branes with $N_f\ll
N_c$ then the open string tachyon is essential for understanding
meson physics, \cite{ckp}. However, the tachyon profile does not
back-react to correct the gauge theory vacuum in this case. A
(closed) string tachyon may be relevant in understanding the vacuum
structure of QCD in the Veneziano limit, $N_{f}\sim N_c$. In this
case it is the avatar of the open string tachyon of the flavor
branes,
 which, because of their large number is becoming effectively a closed string state
(in the sense that it affects the vacuum structure at leading
 order in $1/N_c$).

To summarize, when $N_f=0$ the relevant propagating bulk fields are $g_{\m\n},\phi, a$,
 while there is also a non-propagating $F_5$ field strength.
When $N_c\gg N_{f}\not= 0$, there is also an open string  tachyon $T$
(together with gauge fields on the flavor branes), which is a $N_f\times N_f$
 complex matrix charged under gauge fields of the
$U(N_f)_{L}\times U(N_f)_R$ symmetry, \cite{ckp}. In this paper we will {\em not}
 consider the dynamics of flavor. But we will include
the contribution of the tension of flavor branes in the effective action in this
 section only to indicate the scaling of various terms in
 the effective dilaton potential.

\subsection{The associated branes}

There are several potential branes associated to the various bulk form fields that exist in the theory.
The RR four-form couples to D$_3$ branes, responsible for generating the gauge group.

The RR zero form, dual to the YM $\theta$-angle couples electrically to $D_{-1}$ instantons, which should be  indeed the
YM instantons. It is interesting to mention here that such bulk instanton solutions in the $AdS_5$ case were identified
with the standard YM instanton solutions with the dependence on the radial holographic coordinate playing the role of instanton size.
In our case we expect that small instantons, as they will be close to the AdS$_5$ boundary, resemble almost accurately the
standard YM Instantons. Large instantons however, feel the non-trivial IR geometry and their form is expected to part substantially
from the standard YM form. It may be interesting to study them as they might indicate the correct instanton measure in the IR of YM theory.
This should be the five-dimensional volume form in the non-trivial metric that describes the YM vacuum, together
with a power of the dilaton that needs to be determined. 

The magnetic duals of the $D_{-1}$ instantons are $D_2$ membranes and they couple minimally to the dual $C_3$ form.
They are domain walls in four-dimensions separating different $k$-vacua.
A point-like $D_2$ generates a monodromy for the bulk axion field in the transverse two dimensional space.

There are also point-like $D_0$ branes coupled to the RR one-form $C_1$. They are the baryon vertices which can tie up $N_c$ quarks to form a
large-$N_c$ baryon. Their duals are $D_1$ strings that couple to the dual two-form $C_2$, which represent the flux
 tubes stretched among magnetic quark sources. They will be used in \cite{part2} to investigate the
 interaction between magnetic quarks in confining vacua.

Finally in the NSNS sector we have NS$_0$ branes that are five-dimensional duals of the fundamentals strings.
They are charged magnetically under the NSNS two-form $B_{\m\n}$, and couple directly to the flavor U(1) gauge-fields.
The S-duality map would suggest that they can be viewed as vertices for magnetic baryons.
This is in agreement with the fact that  such magnetic baryons are expected to have masses that scale as ${\cal O}(N_c^2)$ compared to ordinary baryons
whose masses are of order ${\cal O}(N_c)$. The reason is that they contain $N_c$ magnetic quarks, each having a mass of order $N_c$.\footnote{See however a 
slightly different interpretation in \cite{aldo}.}

 \subsection{The effective action}

The (minimal) two-derivative effective action of the perturbative string theory is
   \begin{equation}
S_5=M^3\int d^5x\sqrt{g}\left[e^{-2\phi}\left(R+4(\partial\phi)^2+
{\delta c\over \ls^2}\right)-{1\over 2\cdot 5!}F_5^2-{1\over 2}F_1^2-{N_f\over \ls^2}e^{-\phi}  \right]
 \label{3} \end{equation}
 where as usual
 $F_1=\partial_{\mu}a$
  and we have used the 5-form instead of the zero form. The last term is due to
space filling $D_4-\bar D_4$ brane pairs, with
   $N_f$ is proportional to the number of flavor branes.
  We have explicitly indicated the string scale $\ls$ and the Planck scale
  \begin{equation}
  M^3={1\over g_s^2\ls^3}\sp \dc=10-D=5
 \label{5}  \end{equation}

Anticipating the connection with the gauge theory let us define,
\be\lab{thft} \lambda = N_c e^{\phi}. \ee\ Passing to the Einstein
frame by $g_{\m\n}=\l^{4\over 3}g_{\m\n}^E$ we obtain:
\begin{equation}
S_5=M^3\int d^5x\sqrt{g}\left[N_c^2\le(R-{4\over
3}\frac{(\partial\l)^2}{\l^2}+ {\dc\over \ls^2}\l^{4\over
3}\ri)-{\l^{-{10\over 3}}\over 2\cdot 5!}F_5^2-{\l^2\over
2}(\partial a)^2 -N_c^2{N_f\over N_c\ls^2}\l^{7\over 3} \right].
 \label{6}\end{equation}

We are now at the point where we can solve for the five-form (that controls the number of $D_3$ branes).
Starting from the Einstein frame action
   \begin{equation}
   \delta S_5=-{M^{3}\over 2\cdot 5!}\int d^5 x\sqrt{g}~\l^{-{10\over 3}}~ F_5^2,
    \label{7}\end{equation}
   the equations of motion are $d^*(\l^{-{10\over 3}}F_5)=0$ with the solution
   \begin{equation}
   F_{\mu_1\cdots\mu_5}={N_c\over \ls} \l^{10\over 3}E_{\mu_1\cdots\mu_5}\equiv{N_c\over \ls}
   \l^{10\over 3}{\e_{\mu_1\cdots\mu_5}\over \sqrt{-\det ~g}},
    \label{8}\end{equation}
   where E is the totally antisymmetric {\tt tensor} and $N_c$ is dimensionless and proportional to the number of colors.
   {\tt This is the definition of $N_c$}, and it is obviously unormalized, unless we know explicitly the $D_3$ brane solution of the
    associated string theory.
Inserting back in the equations of motion and  from this reconstructing the action, we obtain
   \begin{equation}
   \delta \tilde S_5=-{M^{3}N_c^2\over 2}\int d^5 x\sqrt{g}~e^{{10\over 3}\phi}
    \label{12}\end{equation} so that
   \begin{equation}
   S_5=M^3N_c^2\int d^5x\sqrt{g}\left[R-{4\over 3}\frac{(\partial\l)^2}{\l^2}-{\l^2\over 2 N_c^2}(\partial a)^2+V(\l)
\right]
    \label{20}\end{equation}
with \be V(\l)= {\l^{4\over 3}\over \ls^2}\left[\dc-x \l-\half
\l^2\right].
 \label{21}\ee
Here we defined the ratio of number of flavors over the number of
colors,
\begin{equation}\label{gam}
    x = \frac{N_f}{N_c}.
\end{equation}
This potential is analyzed in detail in appendix \ref{pertpot}. It
has a minimum at a finite ${\cal O}(1)$ value of the t 'Hooft
coupling\footnote{ We stress again that this is the definition we
use for the 't Hooft coupling and it is unormalized.} \ref{thft}.

This gives rise to an $AdS_5$ solution that was found and analyzed
in \cite{BCCKP}, where it was conjectured to describe the Banks-Zaks
fixed points (for $N_f\sim N_c$, \ie $x\sim 1$). One might think
that a dilaton flow around the $AdS_5$ solution might describe the
gauge theory physics we are seeking. A study of the dilaton
perturbations in appendix \ref{pertpot} shows however that when
$N_{f}\ll N_c$, the dimension of the operator dual to the dilaton
perturbation is well above 6, and this does not compare well with
the dimension-4 perturbation we are looking for. On the other hand,
it is expected that for $N_f\sim N_c$, the solution will be
importantly affected by the tachyon field. Therefore an analysis in
this case that neglects the tachyon effects may be  unreliable. Thus
it is plausible that the $AdS_5$ solution found is relevant for
gauge theories in their Veneziano limit, as argued in \cite{BCCKP}.

Another problem of this AdS vacuum is that, as we show in appendix \ref{pertpot},
 the t 'Hooft coupling at that point is of order one,
and this is not something we expect in the UV region of QCD.

We are therefore led  to the suggestion that the UV of QCD should be described by the $\phi\to -\infty$ asymptotics of the potential
in (\ref{20}). To test this claim, we must study the asymptotic solution of the classical equations and compare it to expectations from QCD.
To do this we make a Poincar\'e-invariant ansatz for the metric. The equations
of motion in the covariant form are,
\begin{equation}
E_{\m\n}-{4\over 3}\left[\pa_{\m}\phi
\pa_{\nu}\phi-{1\over 2}(\pa\phi)^2 g_{\m\n}\right]-{1\over 2}g_{\m\n}V=0\sp
\square\phi+{3\over 8}\frac{dV(\phi)}{d\phi}=0.
    \label{49}\end{equation}
with the Einstein tensor defined as,
\begin{equation}
E_{\m\n}=R_{\m\n}-{1\over 2}Rg_{\m\n},
    \label{50}\end{equation}
We neglect the QCD axion $a$ in our system as it is not expected to affect this solution.\footnote{In
large-$N_c$ theories with strong IR dynamics like QCD,
Witten has argued long time ago \cite{wit1}, that although instanton effects  seem negligible
at large-$N_{c}$, there is non-trivial $\theta$ dependence
in QCD dynamics. This is in particular responsible for the resolution of the $\eta'$ mass
 problem via the Witten-Veneziano mechanism, \cite{wit2}.
Although we expect via such considerations $a$ to have a non-trivial profile in
five dimensions, its ``order of magnitude" is ${\cal O}(1/N_c)$ and therefore
its back-reaction to the other bulk fields can be neglected to leading order \cite{wit3}.}

Let us digress here to introduce the following two coordinate systems we use in the
paper. The ``conformal''
coordinate system is given by,
\begin{equation}
ds^2=e^{2A(r)}\left[dr^2+\eta_{ij}dx^{i}dx^{j}\right],
    \label{conf}
    \end{equation}
and the ``domain wall'' coordinate system is given by,
\begin{equation}
ds^2=e^{2A(u)}\left[\eta_{ij}dx^{i}dx^{j}\right] +  du^2.
    \label{dw}
    \end{equation}
The two are related by\footnote{For simplicity we denote the warp factor in both
  of the coordinate systems by $A$ \ie $A(u) = A(r(u))$.},
\be\lab{radtrans} du = e^{A(r)} dr. \ee The advantage of the
conformal variable $r$ is that in an asymptotically AdS background,
it is directly related to the energy of the gauge theory as $E =
1/r$ in the far UV. Also the corrections to the AdS metric that we describe in
this paper take the simple form of $\log r$. On the other hand, the
domain-wall coordinate $u$ facilitates the solution to the
differential equations and uncovers the relation between the
integration constants and the related gauge theory quantities.
Henceforth, we will denote by
  prime, a
  derivative with respect to $u$ and by dot, a derivative with respect to $r$.

The equations of motion in the conformal coordinate read
\begin{equation}
12\dot{A}^2-{4\over 3}\dot{\phi}^2-e^{2A} V=0 \sp 6\ddot{A}+ 6\dot{A}^2+{4\over 3}\dot{\phi}^2-e^{2A}V=0 \label{54a}\end{equation}
\begin{equation}
\ddot{\phi}+3\dot{A}\dot{\phi}+{3\over 8}e^{2A} \frac{dV}{d\f}=0 \label{55a}\end{equation}
We shall also need the Einstein's equations and
the scalar field equation in the domain-wall coordinate, in the
following form:
\begin{eqnarray}\lab{eins1}
A'^2 = \frac{(\phi')^2}{9} &+& \frac{V}{12}\sp
\lab{eins2}
A'' = -\frac49 (\phi')^2,\\
\phi^{''} &=& -4A' \phi'-\frac38 \frac{d V(\phi)}{d\phi}. \lab{eom2}
\end{eqnarray}

What type of solution do we expect in the UV regime $(r\to 0)$ (or
$u\to -\infty$)? As QCD becomes scale invariant at $E=\infty$, we
expect the space to asymptote to $AdS_5$. Moreover as ${1\over
\lambda}\sim \log E$ we expect that $e^{-\phi}\sim -\log r$.
Therefore, the geometry is expected to be AdS$_5$ modulo logarithmic
corrections. This asymptotic expansion is described in detail in
appendix \ref{afree}.

The potential in (\ref{20}) as $\phi\to -\infty$, behaves as $V\sim e^{a\phi}$ with $a={4\over 3}$.
In  appendix \ref{exponential} we analyze in detail the solutions to the
single exponential potential. We show that in such cases the asymptotics of
the metric is quite different from the expected
behavior outlined above.
Indeed one finds, as $r\to 0$,
\begin{displaymath}
e^{A(r)}\sim \left\{ \begin{array}{ll}
r^{16 \over 9a^2-16} & \textrm{if}\,\, a\ne \frac43\\
e^r & \textrm{if} \,\, a = \frac43.
\end{array}\right.
\label{100}
\end{displaymath}

In fact, the dilaton potential that has the desired behavior in the
UV, must asymptote to a (positive) constant, as $\phi\to -\infty$.
The natural question is whether such a constant can appear from
string theory considerations.

In five dimensions there are higher-order $\a'$ corrections, that contribute to the dilaton potential.
This may sound implausible, as such corrections carry a number of derivatives and
are not therefore part of the potential.
However, it should be remembered that the bulk theory contains the (non-propagating)
 four-form  $C_4$, which is responsible for the charge of $D_3$
branes sourcing the bulk solution. Its field strength is $F_5$.
At the quadratic level, it generates the $e^{{10\over 3}\phi}$ term in the potential, upon dualization. It goes
without saying that its higher-derivative avatars in the effective
action will contribute several more terms in the dilaton potential.
For example the n-th order term (on the sphere) in the $\a'$
expansion is
\be
S_{n}\sim {M^3\over \ell_s^{2-2n}} \int d^5 x
e^{2(n-1)\phi}\sqrt{g}~(F_5)^{2n}
\label{101}\ee
where the dilaton
dependence is dictated by the special transformation properties of
RR field strengths. As explained in detail in appendix \ref{dilpot},
upon dualization, such  terms translate into a dilaton potential of
the form \be V_{F_5}\sim {e^{{4\over 3}\phi}\over
\ell_s^2}\sum_{n=0}^{\infty} a_n ~(N_ce^{\phi})^{2n} \label{102}\ee
where the $a_n$ are dimensionless numbers reflecting the
coefficients of the higher derivative terms like (\ref{101}). Note
that we included the central charge
deficit as the first term in this potential.

This potential is remarkable for two reasons. Firstly, it
is a function of the 't Hooft coupling $\lambda\sim N_c e^{\phi}$, and remains finite in
large-$N_c$ perturbation theory. The second is that it has a regular
expansion for small 't Hooft coupling. On
the other hand it does not contain the term we are looking for,
namely a (positive) constant. In appendix \ref{hd} we indicate how
such an effective term, generating an asymptotic AdS regime can be
generated by higher curvature corrections. We will not however
pursue this line of thought here, instead we will add such a term to
the dilaton potential by {\em fiat}, as was also done in \cite{cr} for example.

The next question to address is what the other terms in the potential should be in order to reproduce the expected logarithmic
running of the gauge theory  coupling constant $\lambda$, that is compatible with QCD perturbation theory.
To answer this question, one must identify the UV energy variable, and this is essentially the coordinate $r$
up to logarithmic corrections.
Matching then the expected equation
\be
{d\lambda\over d\log r}=\beta(\lambda)=-b_0\lambda^2+b_1\lambda^3+b_2\lambda^4+\cdots
\label{103}\ee
with the bulk equations of motion in (\ref{54a}), (\ref{55a}) we obtain that the weak-coupling expansion of the potential should be
\be
V(\lambda)=\sum_{n=0}^{\infty}V_n~\lambda^n
\label{104}\ee
To leading order, $\lambda\sim N_c e^{\phi}$ as can be ascertained by the world-volume
couplings of $D_3$ branes\footnote{If the space has more than five dimensions
or there are other bulk scalar field involved, this identification can change. However this is not the case here.
Another issue involves finite renormalizations of the 't Hooft coupling identification. Such corrections cannot be excluded
but they cannot affect the qualitative analysis, as they involve a constant rescaling of the dilaton. We choose therefore to ignore them. }.
Higher $\alpha'$ corrections involving the five-form on $D_3$ branes can change this identification to
\be
\lambda=\sum_{n=0}^{\infty}c_n ~( N_c e^{\phi})^{(2n+1)}
\label{105}\ee
as is shown in appendix \ref{dilcoup}. In case of ${\cal N}=4$ YM, the leading coefficient is $c_0=4\pi$. However,
in general we consider a possible multiplicative renormalization in the identification of $\lambda$,
hence we keep $c_0$ arbitrary. Moreover, changing $c_0$ is equivalent to changing $b_0$ the first $\beta$-function
coefficient. Therefore we will eventually set $c_0=1$ and we will allow ourselves to vary $b_0$ to reflect this normalization ambiguity.

Such corrections affect the $\beta$-function beyond the first two non-trivial orders, (associated to $b_0,b_1$).
Since higher coefficients are non-universal,
we choose to neglect the renormalizations in (\ref{105}), and set $c_n=0$ for $n\geq 1$.
In the analysis of appendix \ref{afree} we have set $c_0$ to 1. Reinstating it here we obtain
$$
V(\l)={12\over \ell^2}\left[1+{8\over 9}(b_0\l)+{23-36{b_1\over b_0^2}\over 3^4}(b_0\l)^2
-2{324{b_2\over b_0^3}+124+189{b_1\over b_0^2}\over 3^7}(b_0\l)^3+{\cal O}(\l^4)\right]
$$
\be
={12\over \ell^2}\left[1+{8\over 9}(b_0c_0 N_c e^{\phi})+{23-36{b_1\over b_0^2}\over 3^4}(b_0c_0 N_c e^{\phi})^2
-2{324{b_2\over b_0^3}+124+189{b_1\over b_0^2}\over 3^7}(b_0c_0 N_c e^{\phi})^3+\cdots\right]
\label{106}\ee
where $\ell$ is the AdS radius.
We conclude that without knowledge of $c_0$, the effective one-loop term $b^{eff}_0=b_0c_0$ is unknown.

In appendix \ref{afree} it is shown that changing the constant $b_0$ amounts to a redefinition
 of the perturbative QCD scale $\Lambda$ as well as
a redefinition of the non-universal coefficients of the $\beta$-function. This
is nicely reflected in the fact that the dilaton potential (\ref{106}) depends
only on the invariant ratios $b_i/b_0^{i+1}$.
Moreover as shown in the same appendix, the first logarithmic correction to the UV AdS metric is universal, and in particular insensitive to
either $b_0$, or $c_0$.
In view of the above, we will postulate from now on, a dilaton potential with a weak coupling expansion as in (\ref{106}) with $c_0=1$.

\section{Holographic large N$_c$ YM: the vacuum structure\label{gfgt}}

Before we embark on a study of the solution to the equations of motion that describes the vacuum structure of the theory,
we review the action we are going to use
 \be
   S=M^3N_c^2\int d^5x\sqrt{g}\left[R-{4\over 3}{(\partial\l)^2\over \l^2}-
{Z(\l)\over 2N_c^2}(\partial a)^2+V(\lambda)
\right]
    \label{ii6}
\ee
with
\be
V(\lambda)={12\over \ell^2}\left(1+\sum_{n=1}^{\infty}{V_n}\lambda^{n}\right)={12\over \ell^2}\left(1+{8\over 9}b_0\lambda
+{23b_0^2-36b_1\over 81}\lambda^2+\cdots\right)
\label{ii3}\ee
and where we have added a general kinetic function
$Z(\l)$
in front of the axion.
As the axion kinetic energy is suppressed by $1/N_c^2$ it will be neglected in the following investigation of the vacuum solution.
In the action (\ref{ii6}), $N_c$ appears only as an overall coefficient, i.e.
it only changes the definition of the 5-dimensional  Planck scale. Therefore
in the large $N_c$ limit,  quantum corrections can be safely neglected,
as in the standard, critical $AdS/CFT$ correspondence.
All kinetic and potential terms are independent of $N_c$ when written
in terms of $\l$. Therefore, it is useful to define an ``effective'' dilaton
field by
\be
\Phi \equiv \log \l = \phi + \log N_c.
\ee
 From now on, this is the field
we will use, and we will refer to it as the  ``dilaton'' without
any further specification.

\subsection{General geometry and the geometric invariants}

In this section we introduce a rewriting of the equations of motion that facilitates
the construction of  the desired 5D geometry and its connection  to the gauge theory.
We allow for a general dilaton potential $V(\f)$ in (\ref{ii6}) that admits an
expansion of the form (\ref{ii3}) for small $\l$.

We start from the domain-wall parametrization of the metric in
(\ref{dw}). We use the Einstein's equations (\ref{eins1}) and the
scalar field equation (\ref{eom2}). It is a well-known fact that
these second-order equations can be reduced to a system of
first-order equations\footnote{This well-known technique in
differential equations was discussed in this context in
\cite{sken1}.}. This is generally done by introducing a
superpotential\footnote{It goes without saying that this amounts to
choosing one of the two initial conditions for the dilaton.}
 . Here we find it more
convenient to rewrite the equations in terms of the derivative of the superpotential with respect
to the dilaton.
For this purpose, we introduce the {\em phase space variable} $X$:
\begin{equation}
\lab{defx}
X = \frac{\f'}{3A'},
\end{equation}
This variable is useful in our set-up as it is closely related to the gauge theory $\beta$-function.

In terms of $X$, the second order system can be rewritten as the following first-order system
of equations,
\begin{eqnarray}\lab{fprime}
(\f')^2 &=& \frac34 V_0 X^2 e^{-\frac83\int_{-\infty}^{\f} X ~d\f},\\
\lab{Aprime}
(A')^2 &=& \frac{V_0}{12}e^{-\frac83\int_{-\infty}^{\f} X d\f},\\
\lab{pstgen1}
\frac{dX}{d\f}&=&(8X+3\frac{d\log V}{d\f}){(X^2-1)\over 6X}.
\end{eqnarray}
Here, $V_0>0$ is the asymptotic value of the dilaton potential as
$\f\to -\infty$. Given $X$ as a function of $\f$, one solves
(\ref{fprime}) to determine the dilaton and (\ref{Aprime}) to obtain
the geometry. Our strategy in solving the system of equations is
essentially reversing (\ref{pstgen1}) and determining the potential
$V(\f)$, given the function $X(\f)$. In the next subsection, we
explain how to determine $X$, by using the gauge theory input. It is
completely determined by the full $\beta$-function of the gauge theory.
We analyze certain fixed points of (\ref{pstgen1}) in the appendix
\ref{exponential}.

Here, we assume that $X$ is known, and  calculate  the invariants
of the geometry. This is necessary in order to understand the singularity
structure of the solution in the IR.

We solve (\ref{pstgen1}) for $V$ as follows:
\begin{equation}
\lab{potx}
V(\f) = V_0(1-X^2)e^{-\frac83\int_{-\infty}^{\f} X\, d\f},
\end{equation}
where $V_0>0$.

The leading scalar invariants, that are constructed from the Einstein
tensor (defined in (\ref{50})) are,
\begin{eqnarray}\lab{E}
E&\equiv&{E_{\m}}^{\m}=30(A')^2 + 12 A'',\\
\lab{Es}
{E_{\m\n}}E^{\m\n}&=&36\left(5(A')^4+4A'' (A')^2 + (A'')^2\right).
\end{eqnarray}
Using the equation of motion (\ref{eins2}) together with (\ref{fprime}) and (\ref{Aprime}),
we obtain the following results for the first-order invariant:
\begin{equation}\lab{E1}
E = V_0(\frac52-4X^2)e^{-\frac83\int^{\f}_{-\infty} X d\f},
\end{equation}
We show in the next section that the presence of an asymptotic AdS geometry
near the boundary requires $X\to 0$ there. From (\ref{E1})
we observe that $E\to 5V_0/2$ there, as expected from the AdS geometry. This
 provides a consistency check of our formulae.
The second order invariant is,
\begin{equation}\lab{E2}
E\cdot E = \frac{V_0^2}{8}(2-11X^2(1-X^2))e^{-\frac{16}{3}\int_{-\infty}^{\Phi} X d\f}.
\end{equation}

We also present the relation of the phase variable and the superpotential.
In 5D, the superpotential is obtained from the scalar potential by solving,
\begin{equation}\lab{spot}
W^2 -\left(\frac34\right)^2\left(\frac{\6 W}{\6\f}\right)^2
= \left(\frac34\right)^3 V(\f).
\end{equation}
The first order equations that follow from this superpotential are,
\be
A' = -\frac49 W\sp
\f' = \frac{dW}{d\f}\label{seom2}.
\ee
Comparison of
 (\ref{seom2})  with (\ref{Aprime}) and (\ref{fprime})
shows that the superpotential is given as,
\be\lab{spotx}
W = \frac{9\sqrt{V_0}}{8\sqrt{3}} e^{-\frac43\int_{-\infty}^{\Phi} X\, d\f}.
\ee
It is also useful to note that $X$ is proportional to the
derivative of the logarithm of the superpotential:
\be\lab{xspot}
X= -\frac34 \frac{d\log W}{d\f}.
\ee

\subsection{The phase variable $X$ and the $\b$-function}

In this section, we show that given i) the relation between the
't Hooft coupling $\l$ and $\phi$ and ii) the relation between
energy scale of the gauge theory, and the radial variable of the 5D
geometry, one can relate the phase space variable that we introduced
in (\ref{defx}) to the $\beta$-function of the gauge theory.
Both relations receive $\alpha'$ corrections, but we shall
first ignore them and consider their effects in section \ref{apcor}.

To fix the first relation, we consider the metric (\ref{dw}) in the Einstein
frame. There are at least two arguments that lead to the energy-radius dependence:
The first is the Polchinski-Strassler
type of argument \cite{PolStras} that uses the gravitational red-shift to relate
\begin{equation}\lab{redshift}
E = \sqrt{g_{00}}~E_{bulk},
\end{equation}
where the LHS is the energy in 4D field theory measured by an observer
on the boundary of the 5D geometry and the RHS
is the energy of a point-like excitation of the gravitational system.
This gives,
\begin{equation}\lab{errel}
\log(E)\propto A.
\end{equation}
The second argument is to make an infinitesimal shift $u\to
u+\delta$ in the metric and to observe that this shift can be
produced by a dilatation $x^{\mu}\to (1+\rho)x^{\m}$. This
dilatation changes the 4D energy by, $\log(E)\to \log(E) - \rho.$
Equivalently, one makes the above shifts in $u$ and $x$ and demands
that the metric does not change to first order in $\rho$. This
fixes,
$$\rho = - A' \delta$$
hence one arrives at the same result (\ref{errel}).

We note that the relation (\ref{errel}) is the most natural in an
asymptotically AdS geometry where $A\to -u/l$ as $u \to -\infty$
and one identifies $\log E= u $
in the original AdS/CFT correspondence (\cite{peet}). However, we stress
that any function whose leading behavior agrees with (\ref{errel}) in the
UV is a candidate definition for the field theory energy.

Another necessary condition is to have a monotonically decreasing
function of the radial variable $u$. This is necessary in order to
have a well-defined direction for the renormalization group flow.
{}From the Einstein's equation (\ref{eins2}), it is clear that
(\ref{errel}) is guaranteed to satisfy this criterion. This is
because, it follows from (\ref{eins2}) that $A'$ is a monotonically
decreasing function of $u$. Since it is already negative at $u =
-\infty$, $A'$ is forced to be negative in the entire range of $u$.
Thus, $A$ itself should be a monotonically decreasing function of
$u$.

We should also note that the scale factor of the metric in the Einstein frame
is expected to be associated to the C-function of the associated theory.
In particular with our definition
we expect that $E^4\sim e^{4A}$ is proportional to the number of degrees of freedom of the theory with energy up to $E$.

Now, we come to the identification
of the $\b$-function. As discussed in Section \ref{motv},
we identify the 't Hooft coupling as:
\begin{equation}
\lab{gdilrel1}
\l = N_c~e^{\phi} \equiv e^\Phi.
\end{equation}
where we chose to set the coefficient $c_0$ to one here
 by reabsorbing it   in the potential.

On the other hand, the $\beta$-function of the gauge theory is given by,
\begin{equation}
\lab{betagen}
\b(\l) = \frac{d\l}{d\log(E)}=\frac{d\l}{dA}.
\end{equation}
The definition of the phase variable $X$ in (\ref{defx}) together with
the gauge theory-geometry relations (\ref{errel}) and (\ref{gdilrel1})
leads to the following identification of the bulk and boundary quantities:
\begin{equation}\lab{chovst}
X = \frac{\b(\l)}{3\l}.
\end{equation}
In the next section, we shall
make use of (\ref{chovst}) in order to formulate restrictions
on the geometry that follow from the desired properties of
the gauge theory.

\section{Asymptotics of the geometry}
\subsection{Asymptotic freedom and the UV geometry\label{af}}

In an asymptotically free gauge theory, the 't Hooft coupling
behaves as, \be\lab{af1} \l \propto \frac{1}{\log E}, \qquad
\textrm{for large}\,\, E. \ee Using our identification of the energy
scale of the gauge theory with the scale factor of the metric in the
Einstein frame, \ie $\log E \propto A$, we learn that this
requirement means, $\f\propto -\log A$, or in terms of the
phase-space function $X$, \be\lab{UVreq1} X\propto - \frac{1}{A}
\propto -\l. \ee Making use of (\ref{UVreq1}) in (\ref{potx}), one
learns that the geometry dual to an asymptotically free gauge
theory, requires the following form of the scalar potential near
$\l= 0$: \be\lab{af4} V = V_0 + V_1 \l + \cdots \ee as was already
announced in section \ref{motv}. Therefore asymptotic freedom in the UV
requires an asymptotically AdS$_5$ geometry. In addition to this, it
requires an expansion of the form (\ref{af4}) around $\l=0$. In
particular, the presence of the first order term $V_1$ is crucial
for the (leading) logarithmic running of the coupling constant.

One can easily make a consistency check on the condition, (\ref{UVreq1}).
Having an AdS geometry in the UV gives $A \propto u$ near the boundary ($u\to\infty$).
Therefore, near the boundary we have,
\be\lab{af2}
\l \propto \frac{1}{u}.
\ee
Using the eq. (\ref{Aprime}) or (\ref{fprime}) we see that this results in the following
asymptotic form for $X$:
\be\lab{af3}
X \propto \l + \cO(\l^2),\quad {\rm as}\,\,\, \l\to 0.
\ee

In particular, as discussed in section \ref{motv}, the simple set up in (\ref{6})
does not satisfy this asymptotics.
Although we presented ideas on how the two requirements can be reconciled, in this work
 we will assume the presence of the suitable dilaton potential.

\subsection{ The $\a'$ corrections\lab{apcor}}

We now generalize the above discussion by taking into account the
possible $\alpha'$ corrections near the boundary. As is clear from
(\ref{potx}) the requirement  (\ref{af4}) follows from
requiring (\ref{af3}). We would like to determine the asymptotics of
$X$ by taking into account the possible corrections that fall into
the following three classes. First, there is a subclass of the
derivative corrections to the Einstein-Hilbert action studied in
appendix (\ref{dilpot}). Second, there are the derivative
corrections to the probe D$_3$ brane action studied in appendix
(\ref{dilcoup}). The combined effect of these corrections can be
stated as,
\begin{equation}\lab{gst}
\l = \l_s\le(1+c_1\l_s^2+(2a_2c_1+\frac{c_2}{2}+c_1^2)\l_s^4+\cO(\l_s^6)\ri).
\end{equation}
Here,
\begin{equation}\lab{ls}
\l_s = e^{\phi}N_c,
\end{equation}
is a purely bulk quantity, while   we use $\l$ to denote the 't
Hooft coupling of the gauge theory.
The unknown coefficients $a_i$
and $c_i$ are defined in (\ref{c2}) and (\ref{41}).

The third type of correction involves the definition of the gauge
theory energy scale in terms of the metric scale factor. Using the
fact that all of the functions in the geometry can be expressed in
the variable $\l_s$, and that, in the far UV, the radial variable in
AdS exactly corresponds to the energy scale, we may write,
\begin{equation}\lab{enrad}
\frac{d \log E }{dA} = f(\l_s) = 1+ f_1 \l_s^2 + f_2 \l_s^4 +\cdots
\end{equation}
with $f_i>0$, for all $i$. This expansion anticipates that the
variable $\l_s$ asymptotes to zero in the UV, and this will be
checked below. We note that a specific choice for the function
$f(\l_s)$ corresponds to a specific scheme\footnote{The scheme dependence of the renormalization
group running in the holographic setup has been discussed in \cite{sm}.} in computing the
$\beta$-function of the field theory. The reason that the expansion variable
is $\l_s^2$ instead of $\l_s$ is due to the form of  the $\alpha'$-expansion,
 as derived in Appendix \ref{hd}.

Now, we can connect $X$ to $\beta$:
\begin{equation}\lab{bx}
\b = \frac{d\l}{d\log E} =
\frac{d\l}{d\l_s}\frac{d\l_s}{dA}\frac{dA}{d\log E} =
\frac{d\l}{d\l_s} 3X \l_s f(\l_s)^{-1}.
\end{equation}
Eqs. (\ref{gst}) and (\ref{enrad}) give,
\begin{eqnarray}
  X &=& -\frac{b_0}{3}\l+ \frac{b_1}{3}\l^2+ \le(\frac{b_2}{3}-\frac43 c_1b_0+\frac{f_1b_0}{3}\ri)\l^3 +
  \le(\frac{b_3}{3}+\frac43 c_1b_1-\frac{f_1b_1}{3}\ri)\l^4\lab{xex}\\
  {} && +\frac13\le(b_4+b_2(4c_1-f_1)+2c_1f_1b_0+c_1b_0^2-f_1^2b_0+f_2b_0-4a_2c_1b_0-2c_2b_0\ri)\l^5
  +\cdots\nonumber
\end{eqnarray}
We remark that the first two terms are completely determined by the
scheme independent gauge theory input, \ie the first two
$\beta$-function coefficients, $b_0$ and $b_1$\footnote{As
explained earlier, in the correspondence of the bulk and the boundary
quantities, neither $b_0$ nor $b_1$ but only the ratio $b_1/b_0^2$
has an invariant meaning as we do not know the correct
normalization of $\l$}. A very interesting corollary of (\ref{xex})
is that the unknown coefficients of the $\alpha'$ expansion, namely
$a_i$, $c_i$ and $f_i$ are always accompanied by the {\em scheme
dependent} $\beta$-function coefficients $b_2, b_3,\dots$. This, for
example, makes it possible to set all the $f_i$ to zero by choosing
an appropriate scheme in the computation of the full $\beta$-function.
In what follows we shall assume that this scheme is chosen and we
shall set $f_i=0$ for all $i$. We call it the ``holographic scheme".
A final observation in (\ref{xex}) is that the $\alpha'$ corrections
to the effective action discussed in Appendix  \ref{dilpot}, (associated to  $a_i$), enter first at the fourth
order in the expansion.

So far we have dealt with the $\a'$ corrections near the boundary. The
possible corrections in the  IR regime shall be discussed in section \ref{confinement}.

\subsection{The UV geometry}

We can obtain the UV asymptotics of the metric by using
(\ref{xex}) and  the formalism of the previous subsection.
The asymptotic form of the metric to  leading,
subleading and next to subleading order,
turns out to be independent of the corrections discussed above:
\begin{equation}\lab{metUV}
ds^2 = e^{2\frac{u}{\ell}}u^{\frac{8}{9}+2b(\frac49-b)/u+\cdots} dx^2 + du^2,
\end{equation}
where $\ell$ is the radius of AdS$_5$:
\begin{equation}\lab{rads}
\ell^2 = {12\over V_0}.
\end{equation}
We have defined the following ratio of the $\beta$-function coefficients,
\be\lab{betab}
b=\frac{b_1}{b_0^2}.
\ee
This ratio has the interesting property that it is left invariant under a
multiplicative renormalization of the coupling constant. This follows from the
Gell-Mann-Low equation and explained in detail in appendix \ref{afree}.

On the other hand, the asymptotic form of the dilaton to subleading order is,
\be\lab{dilUV}
b_0\l = \frac{\ell}{u} +(b-\frac49)\log u+\cdots
\ee
One can also easily obtain the asymptotic form of the Ricci scalar and find that
$R\to -20/\ell^2$: We have the $AdS_5$ geometry in the UV and
the subleading term in (\ref{metUV}) can be viewed as a perturbation on AdS$_5$.

We also present the same asymptotics in the conformal $r$-coordinate system:
\begin{equation}\lab{metUVr}
ds^2 = \frac{\ell^2}{r^2}\le(1+\frac89 \frac{1}{\log(r\L)}
-\frac89 b \frac{\log(-\log(r\L))}{\log(r\L)^2}+\cdots\ri) \le(dx^idx_i+dr^2\ri)
\end{equation}
\be\lab{dilUVr}
\l = -\frac{1}{b_0\log(r\L)}
+\frac{b}{b_0}\frac{\log\le(-\log(r\L)\ri)}{(\log(r\L))^2}+\cdots
\ee
where $\ell$ is the AdS radius and $\L$ is an integration constant, that
corresponds to the perturbative QCD scale in the dual theory.

We remark that the leading {\em and the subleading} terms in
the metric, (\ref{metUV}), depend neither on the energy-radius coefficients
$f_i$, nor on the particular $\beta$-function coefficients $b_i$.
Therefore the first terms in (\ref{metUV}) are universal and solely follow
from the requirement of asymptotic freedom. However, the subleading term of the metric already introduces
the invariant QCD scale $\Lambda$.
Of course the information on the $\beta$-function
is encoded in the leading term of the dilaton solution.

\subsection{Confinement and the IR geometry\label{confinement}}

Given the geometry, a natural question to ask is whether or not this
geometry describes quark confinement in the IR. On the gauge theory side,
a criterion for confinement is that the Wilson loop operator exhibits
an area law behavior at large distances. This corresponds to an asymptotic linear potential between two
external quarks. In the 5d dual this criterion can be tested
by calculating the classical action of a fundamental string
world-sheet that extends to the Wilson loop at the $AdS_5$  boundary,
as argued in  \cite{malda2, rey}.

A detailed analysis with a classification
of all the confining backgrounds is presented in \cite{part2}.
Here we would like to discuss the qualitative features
of the set-up by focusing on
one class of asymptotics:
\be\lab{xconf2}
X(\l) = -{1\over 2} \left[1 + {a\over \log \lambda}
+ \ldots\right], \qquad \l\to \infty
\ee
The asymptotics of the geometry can be obtained using the definition (\ref{defx}):
\be\lab{detA}
A = A_0 + \frac13 \int_0^{\l}\frac{d\l'}{\l'\, X(\l')},
\ee
and $\f$ from (\ref{fprime}). In the conformal variable, one obtains
the following IR asymptotics (as $r\to\infty$)
\be\lab{air}
A =  - C r^{\a} \ + \ \textrm{subleading},
\ee
where $C$ is an integration constant and we have defined,
\be\lab{alpha}
\a = \frac{3}{3-4a}.
\ee
The dilaton behaves as,
\be\lab{fir}
\f = \frac32 C r^{\a} + \textrm{subleading}
\ee
It is shown in \cite{part2} that these backgrounds confine for $\a\geq 1$.
To see this, one needs to take into account the subleading term in (\ref{fir}).
  In the borderline case, $\a=1$, the geometry is interesting and well-known.
In the string frame the metric asymptotically becomes,
\begin{equation}\lab{stmetIR}
ds^2 \to dx^2 + \frac49 d\f^2, \qquad \f \to \infty.
\end{equation}
This is the linear-dilaton background. The Ricci scalar in the
string frame therefore vanishes in the IR. In the Einstein
frame, the metric has  a singularity at a finite
value  $u=u_0$ (corresponding to $r\to \infty$. There, the scale factor
 $\exp[A]$ shrinks to zero. This is due to the fact that $e^{\Phi}$ and
therefore the gauge coupling diverge at that point. The stringy
corrections to the geometry in this regime can be neglected as in
the string frame the background is known to be exact to all orders
in $\a'$.

The linear dilaton asymptotics are  the borderline of
confining backgrounds. The spectrum of mesons and  glueballs is obtained
by considering the fluctuations of the geometry. Although
exhibiting a mass gap, the spectrum  is partly
continuous  for $\a=1$. Therefore, we discard the
borderline case $\a=1$ ($a=0$) for physical  reasons. In the case
$\a>1$ however, the spectrum turns out to be discrete.

For $\alpha>1$ (or $a>0$) the linear dilaton background
in (\ref{stmetIR})
is modified into the following asymptotic metric:
\be\lab{conff}
ds^2 \to  \f^{\frac{4a}{3}}dx^2 + \frac49 \f^{-\frac{4a}{3}} d\f^2, \qquad \f \to \infty.
\ee
The Ricci scalar in the string frame reads asymptotically ,
\be\lab{riccistf}
R_s  \sim -4 a (7 a-3) \f^{\frac43 a - 2}.
\ee
We observe that for the confining backgrounds $\alpha>1$
(or $0<a<3/4$) the string frame Ricci scalar vanishes in the IR. This implies that
the singularity in the string frame is only due to the diverging dilaton.
This is a good type of singularity where the string moving in this background
feels very small curvature corrections.

The effect of $\a'$ corrections  to the confinement criteria is an
important question to which we turn.
Fortunately, as happens with ``good" holographic singularities, the
low-lying fluctuations have an effective potential
that shields them from the singularity.
Therefore, the $\a'$ corrections are under control in this case, see \cite{part2}.

Let us finally note the behavior of the dilaton potential and the
various geometric invariants in this region. The dilaton potential
follows from (\ref{potx}): \be\lab{potIR} V \approx \frac34 V_0
\l^{\frac43}(\log\l)^{\frac{\a-1}{\a}} \le(1+
\co(\frac{1}{\log\l})\ri). \ee Similarly one obtains the following
asymptotics for the geometric invariants in the Einstein frame,
\be
\lab{geoinvIR} E \sim R
\sim (\6\f)^2 \sim \l^{\frac43}(\log\l)^{\frac{\a-1}{\a}}.
\ee

We see that this is the linear dilaton background with logarithmic
corrections. In the $\a=1$ case the leading behavior of the geometry
is exactly that of the linear dilaton background.

\section{Examples of background geometries}

In this section we present  examples of full background geometries that exhibit the desired
UV and IR asymptotics  discussed in the previous two sections. The first is an example showing asymptotic freedom in the UV and confinement
in the IR. For comparison, we present an example which is asymptotically free in the UV, but has a conformal fixed point in the IR.
Finally, in appendix \ref{bapp}  we present a third example which is asymptotically free in the UV and has
a $\beta$-function that asymptotes to zero in the IR
with an exponential tail.

\subsection{Standard QCD type two-loop $\beta$-function  }
\label{twoloop}

We start from  the ``exact" $\beta$-function: \be\lab{tlbeta} \b(\l)
= -\frac{3b_0\l^2}{3+2b_0\l}-\frac{(2b_0^2+3b_1)\l^3}
{3(1+\l^2)\left[1+{(2b_0^2+3b_1)\over 18a}\log(1+\l^2)\right]}. \ee
with $b_{0,1}>0$, $a>0$. This expression is chosen so that one has the desired asymptotics
 in the UV and IR and such that there are no poles or branch cuts in $\l$. In particular at weak
coupling it reproduces the QCD result to two loops, $\b=-b_0 \l^2 -
b_1\l^3+{\cal O}(\l^4)$. At strong coupling it behaves as
$\beta=-{3\over 2}\lambda+3a{\lambda\over \log \lambda}+\cdots$ as required by  (\ref{chovst}) and (\ref{xconf2}).
 It also avoids unwanted singularities or branch points
at finite values of $\l$.

\begin{figure}
 \begin{center}
 \leavevmode \epsfxsize=12cm \epsffile{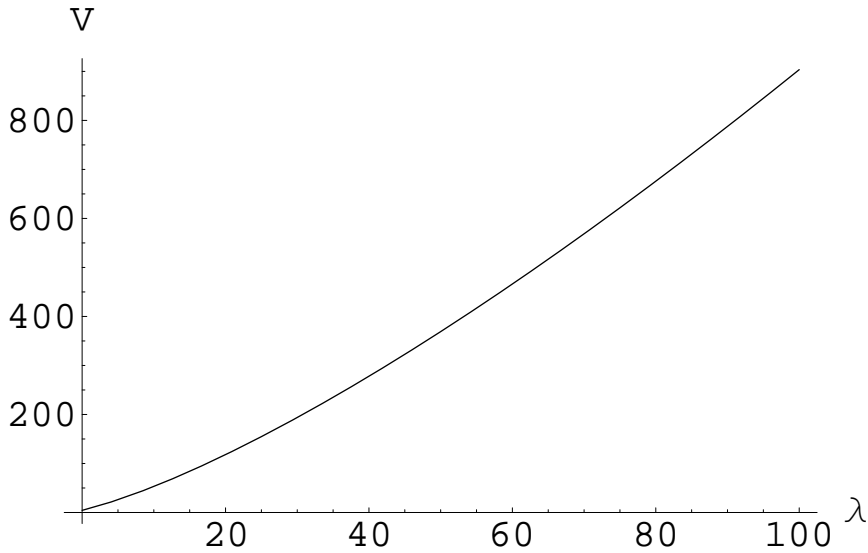}
 \end{center}
 \caption[]{Dilaton potential as a function of $\l$ for the solution studied in section \ref{twoloop}. }
 \label{fig1}\end{figure}

\begin{figure}
 \begin{center}
 \leavevmode \epsfxsize=12cm \epsffile{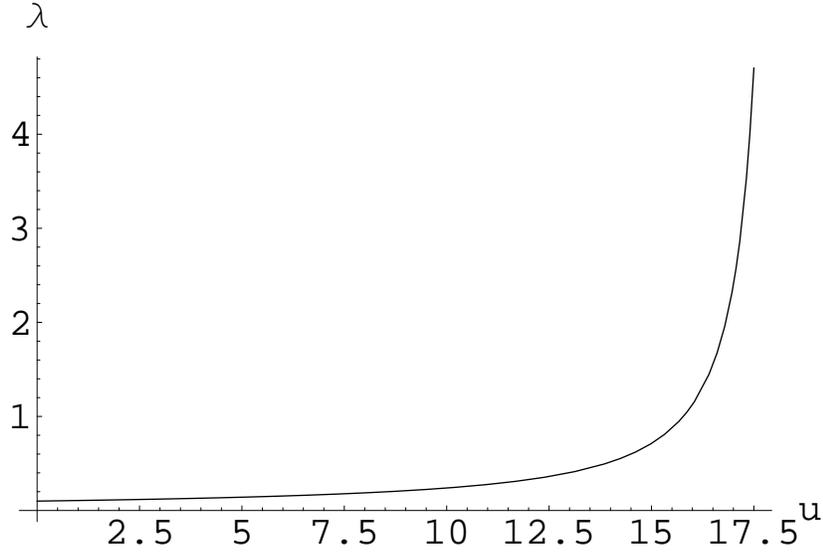}
 \end{center}
 \caption[]{The running of the coupling constant. It diverges at $u_0$. }
 \label{fig2}\end{figure}

According to (\ref{potx}) the dilaton potential that corresponds to this specific $\beta$-function is,
\begin{equation}\lab{pot1e}
V(\l) = V_0
(1-X(\l)^2)(3+2b_0\l)^{\frac43}\le(18a+(2b_0^2+3b_1)\log(1+\l^2)\ri)^{\frac{8a}{3}},
\end{equation}
where the phase-space variable $X(\l)$ is given by $\b(\l)/3\l$. The
equation that determines the running of $\l$ follows from,
(\ref{fprime}):
\begin{equation}\lab{l4e}
\l' = \frac{3}{\ell} X(\l)
(3+2b_0)^{\frac23}\le(18a+(2b_0^2+3b_1)\log(1+\l^2)\ri)^{\frac{4a}{3}}.
\end{equation}
We give the plot of the potential in fig \ref{fig1}. The solution of
(\ref{l4e}) as a function of $u$, with the initial condition
$\l(0)=0$ is shown in fig \ref{fig2}.

We see from the figures that $\l$ diverges in the IR at the
singularity $u_0$.

The associated geometry can be derived as follows. The metric in
the string frame is,
\begin{equation}\lab{mets1}
ds^2  = e^{2A+\frac43\f}dx^2+ e^{\frac43\f}dr^2 = \l^{\frac43}(e^{2A}dx^2 + (\l')^{-2} d\l^2).
\end{equation}
The metric function $A$ is obtained as a function of $\l$ from
(\ref{detA}). Unfortunately, it is not possible to express this
integral in terms of elementary functions. The Einstein frame metric
is obtained from (\ref{mets1}) by multiplying with $\l^{-4/3}$. The
asymptotics of this geometry are discussed in the previous section.

One can also easily determine certain invariants of the geometry.
The Ricci scalar is given by (\ref{E})
and $R = -3/2 E$. In the Einstein frame, one obtains,
\begin{equation}\lab{r1}
  R = -\frac32 V_0 (\frac52-4X(\l)^2)(3+2b_0\l)^{\frac43}
\le(18a+(2b_0^2+3b_1)\log(1+\l^2)\ri)^{\frac{8a}{3}}.
\end{equation}
The asymptotic behavior of the various geometric invariants in this geometry
are also presented in the previous two sections.

\subsection{$\beta$-function with an IR fixed point  }
\label{bankszaks}

We choose an exact $\beta$-function as
\begin{equation}\lab{x2e}
\b(\l)= \frac{-b_0\l^2 +\bb_1\l^3}{1+\frac{2b_0}{3}\l-\frac{2\bb_1}{3}\l^2},
\end{equation}
with
\be\lab{bb1}
\bb_1 = b_1-2b_0^2/3>0.
\ee
In perturbation theory we have $\b=-b_0 \l^2+b_1\l^3 +{\cal O}(\l^4)$ with $b_{0,1}>0$.

According to (\ref{potx}) the potential for this specific $\beta$-function is,
\begin{equation}\lab{Apot1e}
V(\l) =
\frac{V_0}{(\l_+)^{a_+}(-\l_-)^{a_-}}\le((\l_+-\l)^2(\l-\l_-)^2-\frac14
\l^2(\l_0-\l)^2\ri)(\l_+-\l)^{a_+-2}(\l-\l_-)^{a_--2}.
\end{equation}
Here we defined the following quantities. $\l_0=b_0/\bb_1$ is the
value of the coupling constant at the IR fixed point. $\l_{\pm}$ is
defined as:
\begin{equation}\lab{lampm}
\l_{\pm}= \frac{\l_0}{2}\pm\frac{\sqrt{\l_0^2+\frac{6}{\bb_1}}}{2},
\end{equation}
and the exponents are,
\begin{equation}\lab{aapm}
a_{\pm}= -\mp\frac43\frac{\l_{\mp}}{\l_+-\l_-}.
\end{equation}
We note that along the RG flow, the range of $\l$ is such that,
$\l_-<0<\l<\l_0<\l_+$.

\begin{figure}
 \begin{center}
 \leavevmode \epsfxsize=12cm \epsffile{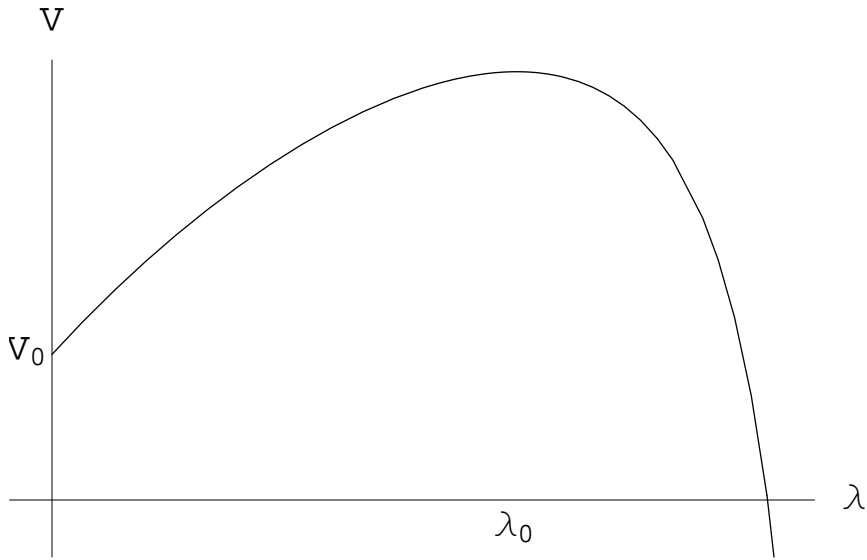}
 \end{center}
 \caption[]{Dilaton potential as a function of $\l$ when there is an IR fixed point.
We have set $V_0=b_0=b_1=1$ for simplicity. }
 \label{fig3}\end{figure}

We give the plot of the potential in fig. \ref{fig3}.
The RG flow is from
$\l=0$ in the UV toward $\l=\l_0$ in the IR. As we discussed in
section 4 the UV geometry is AdS with the radius given by eq.
(\ref{rads}). As fig. \ref{fig3} suggests, the IR end of the potential is
also AdS.

Below we present the details of the IR geometry.
The IR AdS radius is obtained by taking the $\l\to\l_0$ limit in the
potential, (\ref{Apot1e}):
\begin{equation}\label{IRrads}
    \ell_{IR} = 2\sqrt{\frac{3}{V_f}}, \qquad V_f = V_0
    \le(\frac{-\l_-}{\l_+}\ri )^{a_+-a_-}.
\end{equation}
We may deduce from (\ref{IRrads}) that $V_f/V_0>1$. Therefore, the c-theorem of \cite{freedmangubser}
is obeyed for this holographic RG flow.

\begin{figure}
 \begin{center}
 \leavevmode \epsfxsize=12cm \epsffile{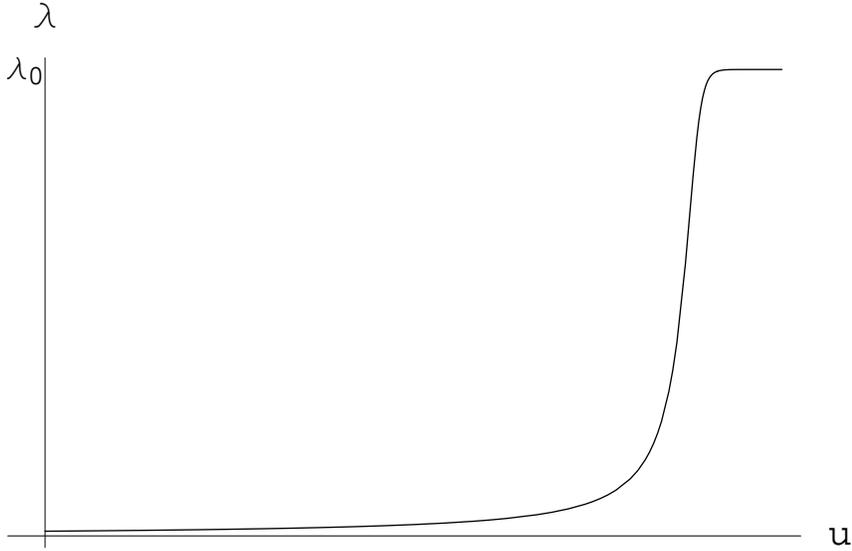}
 \end{center}
 \caption[]{Running of the coupling constant in the Banks-Zaks case.
We have set $V_0=b_0=b_1=1$.}
 \label{fig4}\end{figure}

The equation that determines the running of $\l$ follows from,
(\ref{fprime}) as,
\begin{equation}\lab{l2e}
\l' =
\frac{(3V_0)^{\half}}{2}\l^2(\l_0-\l)(\l_+-\l)^{\frac{a_+}{2}-1}(\l-\l_-)^{\frac{a_-}{2}-1}.
\end{equation}
The solution of (\ref{l2e}) as a function of $u$,
with the initial condition $\l(0)=0$ is shown in fig \ref{fig4}. We observe that, $\l$ flows into a non-trivial
fixed point at low energies. The value of the coupling constant at the fixed point is $\l_0=b_0/\bb_1$.
As $\l$ increases from 0 to $\l_0$ during the RG flow, one has the inequality $\l_-<0<\l<\l_0<\l_+$
in this range.

 The metric in
the string frame (\ref{mets1}) is obtained by using (\ref{detA}) and (\ref{fprime}) as,
\begin{equation}\lab{dmets3}
ds^2  = e^{2A_0+\frac{2}{b_0\l}}\l^{-\frac{2\bb_1}{b_0^2}}(\l_0-\l)^{\frac{2\bb_1}{b_0^2}}dx^2
+ \frac{16}{V_0}(\l_0-\l)^{-2}(\l_+-\l)^{-a_+}(\l-\l_-)^{-a_-}\frac{d\l^2}{\l^{\frac83}}.
\end{equation}

One can also easily determine certain invariants of the geometry.
The Ricci scalar is given by (\ref{E})
and $R = -3/2 E$. In the Einstein frame we obtain,
\begin{equation}\lab{r3}
  R = -\frac34 V_0 \le(5(\l-\l_-)^2(\l_+-\l)^2-2\l^2(\l_0-\l)^2\ri)(\l-\l_-)^{a_-}(\l_+-\l)^{a_+}
\end{equation}
Note that $R$ goes to a constant in the IR, \ie as $\l\to\l_0$. This is in accordance with
the presence of the fixed point in the IR.
Similarly the invariant $(\6\f)^2$ follows from (\ref{dmets3}) as,
\begin{equation}\lab{df3}
(\6\f)^2 = g^{\l\l}(\6_{\l}\f)^2 = \frac{V_0}{16}\l^2(\l-\l_-)^{a_-}(\l_+-\l)^{a_+}(\l_0-\l)^2,
\end{equation}
which vanishes in the IR as required from the conformality.

The IR asymptotics of the geometry can be obtained by solving the
equations (\ref{Aprime}) and (\ref{fprime}) near $\l_0$ (that
corresponds to $u=\infty$). We present the geometry in the conformal
coordinate, by performing the change of variable (\ref{radtrans}).
One finds,
\begin{equation}\label{BFIR}
    ds^2 =
    \frac{l_{IR}^2}{r^2}(1+\frac{1}{r\Lambda})^{-2}\le(dx^2+dr^2\ri),
\end{equation}
near $r\to\infty$. Here $\Lambda$ is an integration constant of
equation (\ref{fprime}). The gauge theory interpretation of
$\Lambda$ is the energy scale below which the conformality in the IR sets
in.

We may determine the exact form of the quark potential and
check that it indeed has the conformal form in the IR. We find the
following relation between the quark-anti-quark potential and the
distance in the IR asymptotics,
\begin{equation}\lab{qpot2}
E = \frac{C[\l_0,\bb_1]^2}{L},
\end{equation}
with the constant $C$ is given by,
\begin{equation}\lab{constC}
C[\l_0,b_1] = \frac{2\bb_1^{-\half}}{\sqrt{3V_0}}\frac{\l_0^{\frac53}}
{(\l_+-\l_0)^{\frac{a_+}{2}}(\l_0-\l_-)^{\frac{a_-}{2}}}.
\end{equation}
This exhibits conformal behavior.

Of course, as the IR theory is conformal, we do not expect this
model to be a good for description of pure YM type theories.
We have presented it however, for illustration purposes.

\section{Fluctuations near the boundary}

The general geometry discussed in the previous section is asymptotically
AdS near the boundary, $u\to\infty$.
It is important to study the fluctuations
of the fields on this background. According to the standard rules of AdS/CFT,
near the boundary, these fluctuations should correspond either to sources or
expectation values of the corresponding operators of the gauge theory. In what
follows, we compute the UV asymptotics of the fluctuations keeping the
subleading terms as well. According to the principles of AdS/CFT these
subleading terms should give information on the anomalous dimensions of the
operators dual to the fluctuations.

{}For the case of a 5D geometry coupled to one scalar field the analysis of the fluctuations
have been carried out in \cite{bianchi,Kofman} (see also \cite{hol} and \cite{KN}).
We are interested in the fluctuations of the metric, the dilaton and the axion.
The metric and the dilaton can be decomposed into 4D traceless-transverse
tensor $h_{\m\n}$, two transverse gauge fields $A_{\m}$ and $V_{\m}$ and 5 scalar fields.

The tensor modes  obeys the usual 5D wave equation:
\begin{equation}\lab{tensf}
\Box_5 h_{\m\n} = 0,
\end{equation}
where $\Box_5$ is the scalar Laplacian on the 5D space-time\footnote{In this section
  we use the conformal coordinate system, eq. (\ref{conf})}:
\begin{equation}\lab{5Dlap}
\Box_5 = e^{-2A}\le(\frac{d^2}{dr^2}+3\frac{dA}{dr}\frac{d}{dr}+\Box_4\ri).
\end{equation}
{}From (\ref{tensf}), it is clear that the non-normalizable solution
for $h_{\m\n}$ sources a dimension 4 operator. It is the
energy-momentum tensor of the 4D boundary theory. Also it can be
seen that the vector fluctuations can completely be gauged away by
using 5D diffeomorphisms.\footnote{This is strictly true only for the 4D massive modes, \cite{KN},
which is the case we are interested in.} Thus, in what follows we focus on the
interesting case of the scalar fluctuations.

\subsection{Dilaton fluctuations}

As discussed in detail e.g. in \cite{KN}, there is
a single diffeomorphism  invariant combination among the 5 scalar fluctuations:
\begin{equation}\lab{xi}
\zeta = \psi -\frac{1}{3X} \k.
\end{equation}
Here $X$ is the variable defined in (\ref{defx}), $\psi$ is related to the trace of
the metric fluctuation and $\chi$ is the fluctuation of the dilaton.
We note from (\ref{UVreq1}) that, for any asymptotically-free boundary theory,
$X$ approaches zero in the UV. As a result, close to the boundary,
$\psi$ and $\chi$ decouple from each other and $\zeta$ reduces to
the dilaton fluctuation. In the conformal coordinate system (\ref{conf}),
it obeys the following equation:
\begin{equation}\lab{scaf}
\ddot{\zeta}+(3\dot{A}+2\frac{\dot{X}}{X})\dot{\zeta}- k^2 \zeta = 0,
\end{equation}
where $k^{\mu}$ is the 4D momentum.

It is interesting to comment on the correspondence of the gauge invariant
scalar fluctuation $\zeta$, with the Yang-Mills operators.
Indeed, it corresponds to the ``dressed" $Tr[F^2]$  operator
\be
\zeta~~\leftrightarrow ~~\beta(\lambda)Tr[F^2]
\ee
which is known to be RG invariant (see for example \cite{narison}).
The  invariance of $\zeta$ under infinitesimal $r$ reparametrizations,
 guarantees that it corresponds to a RG-invariant operator.

Using the asymptotic forms of the dilaton and the metric in
(\ref{metUVr}) and (\ref{dilUVr}) close to the boundary,
(\ref{scaf}) becomes (in  units where $\ell=1$),
\begin{equation}\lab{scaf1}
\ddot{\zeta} -\frac1r\le(3+\frac{2}{\log r\L}+\frac{C}{\log^2 r\L}
-2b\frac{\log(-\log r\L)}{\log^2 r\L}\ri)\dot{\zeta}-k^2\zeta= 0,
\end{equation}
where $b$ is defined in (\ref{betab}) and
the coefficient $C$ is given by,
\be\lab{coefC}
C = \frac43+2b(2-\log b_0).
\ee
The corresponding fluctuation equation in the pure AdS$_5$ geometry is given by,
\begin{equation}\lab{scafads}
\ddot{\zeta} -\frac3r\dot{\zeta}-k^2\zeta= 0,
\end{equation}

There are two independent solutions to (\ref{scaf1}), one is normalizable and
the other is non-normalizable. The asymptotic expansion of the
non-normalizable solution reads,
\bea\lab{xin}
\zeta_{source} &=& 1 - \frac{(kr)^2}{4}\le[1-\frac{1}{\log
  r\L}+\frac{2-C}{2\log^2r\L}+b\frac{\log(-\log
  r\L)}{\log^2r\L}+\cdots\ri]\nonumber\\
{}&& + \frac{(kr)^4}{64}\le[1-\frac{3}{2\log r\L}
+\frac{7-3C}{4\log^2r\L}+\frac{3b\log(-\log r\L)}{2\log^2r\L}+\cdots\ri]+\co(r^6).
\eea
According to the standard rules of the AdS/CFT correspondence this solution is
dual to a {\em source term} for a gauge-invariant operator of engineering dimension
4 and parity +1 in the boundary gauge theory. As argued above such an operator is
$\beta(\l)~\tr F^2$.

The other solution of (\ref{scaf1}) is normalizable near the boundary and it
has the following asymptotic expansion:
\be\lab{xinn}
\zeta_{vev} = (kr)^4\le[\log^2r\L+2b\log r\L\,
\log(-\log r\L)+\le(2b-C-\half\ri)\log r\L+\cdots\ri]+\co(r^6)
\ee
This solution corresponds to a VEV for $\tr F^2$, namely a gluon condensate.

\subsection{Axion fluctuations}

The other scalar in the geometry is the axion. Its background value
is non-zero if $\theta_{UV}\not= 0$ and is obtained by solving
\be
{1\over \sqrt{g}}\partial_{\m}\left[Z(\Phi)\sqrt{g}g^{\m\n}\partial_{\nu} \right]a=0
\ee
in the Einstein frame.
$Z$ is an ``effective" potential that resums higher-order corrections due to the five form, as explained in appendix
\ref{axion-corr}.
The background solution $a(r)$ describes the RG running of the QCD $\theta$ angle.
As the axion is suppressed by extra powers of $1/N_c$, \cite{wit3}, this equation can be solved in the background already
determined by the metric-dilaton equations.
We analyze it in detail in \cite{part2} where the vacuum condensate of $Tr[F\wedge F]$ is calculated.
Here we note that since the action for the axion is quadratic, its background value does not affect its fluctuations
that are dual to the $Tr[F\wedge F]$ operator in QCD.

The equation for the axion fluctuations follow from (\ref{6}) and
we obtain ,
\begin{equation}\lab{axf}
\ddot{\a}+\left(3\dot{A}+2{\dot Z\over Z}\right)\dot{\a}- k^2 \a = 0.
\end{equation}

The non-normalizable axion fluctuation corresponds to a source term for the
operator $\tr F\wedge F$ and the normalizable solution is dual to a VEV
for the same operator, $\langle \tr F\wedge F\rangle$,
namely an axial glueball condensate. This will be further determined in \cite{part2}.

 \addcontentsline{toc}{section}{Acknowledgments}
\acknowledgments

\noindent It is a pleasure to thank G. Bali, B. Bringoltz, R. Casero,  L. Giusti, D. K. Hong,
K. Intriligator, M. Luscher, J. Mas, H.B. Meyer,  C. Morningstar, V. Niarchos,
C. Nunez, H. Panagopoulos, I. Papadimitriou, S. Pal,  A. Paredes, G. Poliscastro, F. Sannino,
C. Skenderis, M. Shifman, E. Shuryak, S. J. Sin, J. Sonnenschein, M. Teper, J. Troost,
A. Vainshtein,
G. Veneziano, A. Vladikas,  and especially  F. Nitti
for useful discussions and insights.
UG is supported by European Commission Marie Curie
Postdoctoral Fellowships, under contract MEIF-CT-2006-039962.
This work was also partially supported by
INTAS grant, 03-51-6346, RTN contracts MRTN-CT-2004-005104 and
MRTN-CT-2004-503369, CNRS PICS \#~3059 and 3747 ,
 and by a European Union Excellence Grant,
MEXT-CT-2003-509661.

\newpage

\newpage
\appendix
\renewcommand{\theequation}{\thesection.\arabic{equation}}
\addcontentsline{toc}{section}{Appendices}
\section*{APPENDIX}

\section{Asymptotic expansions in the (UV) weak-coupling regime\label{afree}}
\def\a{\alpha_{s}}
In this appendix we will carefully analyze the weak coupling
expansion in the UV regime of the space, near the boundary. We start
from the equations of motion in the conformal frame as described in
section \ref{motv},
\begin{equation}
ds^2=e^{2A(r)}\left[dr^2+\eta_{ij}dx^{i}dx^{j}\right]\sp E_{rr}=6\dot A^2\sp
E_{ij}=3\left[\ddot A+\dot A^2\right]\eta_{ij}
    \label{52}
    \end{equation}
where we use the mostly plus convention for the metric.
The two scalar invariants of the metric up to quadratic order are
\begin{equation}
{E_{\m}}^{\m}=-{3\over 2}R={6 e^{-2A}}\left[2\ddot A+3 \dot A^2\right]\sp
{E_{\m\n}}E^{\m\n}={36~e^{-4A}}\left[\ddot A^2+2\ddot A\dot A^2+2\dot A^4\right]
    \label{53}\end{equation}

In this ansatz the equations (\ref{49}) become
\begin{equation}
12\dot A^2-{4\over 3}\dot\phi^2-e^{2A} V=0 \sp 6\ddot A+6\dot A^2+{4\over 3}\dot\phi^2-e^{2A}V=0 \label{54}\end{equation}
\begin{equation}
\ddot\phi+3\dot A\dot\phi+{3\over 8}e^{2A} V'=0\sp V'={dV(\phi)\over d\phi} \label{55}\end{equation}

To continue further we define the 't Hooft coupling and its inverse as usual
\begin{equation}
\l=N_c e^{\phi}\sp \a={1\over \l}\sp \dot\phi=-{\dot\a\over \a}\sp
\ddot\phi=-{\ddot\a\over \a}+{\dot\a^2\over \a^2} \label{56}\end{equation}
(\ref{54}) and (\ref{55}) become
\begin{equation}
-{\ddot\a\over \a}+{\dot\a^2\over \a^2}-3\dot A{\dot\a\over \a}+{3\over
8}e^{2A} V'=0
\sp
12\dot A^2-{4\over 3}{\dot\a^2\over \a^2}-e^{2A}V=0
\label{58}\end{equation}
 We now change variables to
\begin{equation}
r=e^{-t}\sp e^{A(t)}=e^t d(t)\sp \label{59}\end{equation} where $t$ is
essentially the log of the energy in the perturbative region. The equations become
\begin{equation}
-{\a''\over \a}-4{ \a'\over \a}+{\a'^2\over \a^2}-3{
\a'\over \a}{ d'\over d}+{3\over 8}d^2 V'=0
\label{60}\end{equation}
\begin{equation}
12\left(1+{ d'\over d}\right)^2-{4\over 3}{ \a'^2\over
\a^2}-d^2V=0 \label{61}\end{equation}
where we now use primes for $t$ derivatives.

We assume a regular power expansion for the potential
\begin{equation}
V=V_0+V_1\l+V_2\l^2+V_3\l^3+V_4\l^4+\cdots\sp
V'=V_1\l+2V_2\l^2+3V_3\l^3+4V_4\l^4+\cdots \label{62}\end{equation}

We now substitute for the inverse coupling
\begin{equation}
{1\over \l}=\a=L-{b_1\over b_0}\log L+{b_1^2\over b_0^2}{\log L\over
L}+ \left({b_1^2\over b_0^2}+{b_2\over b_0}\right){1\over L}+
{b_1^3\over 2b_0^3}{\log^2 L\over L^2}+ \label{63}\end{equation}
$$+
{b_1b_2\over b_0^2}{\log L\over L^2}+
\left[{b_3\over 2b_0}-{b_1^3\over 2b_0^3}\right]{1\over L^2}+
{\cal O}\left({1\over L^3}\right)
$$
\begin{equation}
L=a_0+b_0 t=-b_0\log(r\Lambda)\sp L'=b_0
\label{64}\end{equation}
The expression above is taylored so that coupling satisfies the standard RG
equation
\begin{equation}
\lambda'=-b_0\l^2+b_1\l^3+b_2\l^4+b_3\l^5\cdots\sp \a'=b_0-{b_1\over \a}-{b_2\over \a^2}-{b_3\over \a^3}+\cdots
\label{65}
\end{equation}
where $\Lambda$ is the RG invariant scale of QCD.

We may rewrite now equations (\ref{60}), (\ref{61}) as
\begin{equation}
{d'\over d}={1\over 8}{\a\over  \a'}d^2V'-{4\over 3}-{1\over
3}{\a''\over  \a'}+{1\over 3}{\a'\over \a}
\sp
{d'\over d}=-1+\sqrt{{d^2V\over 12}+{1\over 9}{\a'^2\over
\a^2}} \label{70}\end{equation}
so that
\begin{equation}
{1\over 8}{\a\over  \a'}d^2V'-{1\over 3}-{1\over 3}{\a''\over
 \a'}+{1\over 3}{\a'\over \a}= \sqrt{{d^2V\over 12}+{1\over
9}{ \a'^2\over \a^2}} \label{71}
\end{equation}
Note that only one
branch of the square root is relevant. We expand the potential in inverse powers of logs to obtain
\footnote{We thank Shesansu Pal for pointing out errors in the following two equations
in the first version of this paper.}
\begin{equation}
V=V_0+{V_1\over L}+{V_2\over L^2}+{b_1\over b_0}V_1{\log L\over
L^2}+\left(V_3-{b_2\over b_0}V_1-{b_1^2\over b_0^2}V_1\right){1\over
L^3}+ \label{73}\end{equation}
$$+{b_1^2\over b_0^2}V_1{(\log L)^2\over L^3}+\left(2{b_1\over b_0}V_2-{b_1^2
\over b_0^2}V_1\right){\log L\over L^3}+{\cal O}\left({1\over L^4}\right)
$$

\begin{equation}
{\a\over  \a'}V'={V_1\over b_0}+\left({2V_2\over
b_0}+{b_1V_1\over b_0^2}\right){1\over L} +\left({3V_3\over
b_0}+{b_2V_1\over b_0^2}\right){1\over L^2}+ \left({2b_1V_2\over
b_0^2}+{b_1^2V_1\over b_0^3}\right){1+\log L\over L^2}+{\cal
O}\left({1\over L^3}\right) \label{75}
\end{equation}
{}From (\ref{71}) we obtain
\begin{equation}
{3\over 8}d^2={V+{\a\over \a'}V'\left(1+{\a''\over
\a'}-{\a'\over \a}\right)\over \left({\a\over \a'}V'\right)^2}\pm
\ee
\be
\pm {\sqrt{ \left[V+{\a\over
\a'}V'\left(1+{\a''\over \a'}-{\a'\over
\a}\right)\right]^2- \left(1+{\a''\over \a'}-2{\a'\over
\a}\right)\left(1+{\a''\over \a'}\right)\left({\a\over
\a'}V'\right)^2}\over \left({\a\over \a'}V'\right)^2}
\label{77}\end{equation}

Using
\begin{equation}
V+{\a\over \a'}V'\left(1+{\a''\over \a'}-{\a'\over
\a}\right)= V_0+{V_1\over b_0}+\left({2V_2\over b_0}+{b_1V_1\over
b_0^2}\right){1\over L}+ \left({2b_1V_2\over b_0^2}+{b_1^2V_1\over
b_0^3}\right){\log L\over L^2} \label{78}\end{equation}
$$
+\left({3V_3\over b_0}+{2b_1V_2\over b_0^2}-V_2-{b_2V_1\over b_0^2}+
{b_1^2V_1\over b_0^3}\right){1\over L^2}+{\cal O}\left({1\over L^3}\right)
$$
we obtain
\begin{equation}
{3\over 8}d^2={b_0\over V_1}\left(1+{b_0V_0\over
V_1}\right)-{(2b_0V_0+V_1)(b_1V_1+2b_0V_2)\over V_1^3 L}
-{b_1(2b_0V_0+V_1)(b_1V_1+2b_0V_2)\log L\over b_0V_1^3 L^2}
\label{79}\end{equation}
$$
+\left[-{3b_0(2b_0V_0+V_1)V_3\over V_1^3 L^2}+{4b_0(3b_0V_0+V_1)V_2^2\over V_1^4L^2}-{b_0^2V_2\over V_1^2 L^2}
+{2b_1(4b_0V_0+V_1)V_2\over V_1^3L^2}\right.
$$
$$\left.
+{b_2(2b_0V_0+V_1)\over V_1^2L^2}+{b_1(b_1V_0+b_0V_1)\over V_1^2L^2}\right]\pm
$$
$$
\pm\left[{b_0^2\over V_1^2}\sqrt{V_0\left(V_0+2{V_1\over b_0}\right)}
-\sqrt{b_0}{(2b_0b_1V_0^2V_1+3b_1V_0V_1^2-b_0V_1^3+4b_0^2V_0^2V_2+6b_0V_0V_1V_2)\over V_1^3\sqrt{V_0(2V_1+b_0V_0)}L}
+{\cal O}\left({1\over L^2}\right)\right]
$$

Compatibility to leading order with the first of (\ref{70}) implies
\begin{equation}
{V_1}={8\over 9}b_0V_0 \label{80}\end{equation} for the plus sign.
The minus sign does not lead to a consistent solution. Therefore to
 leading order
\begin{equation}
{3\over 8}d^2V_0={b_0V_0\over V_1}\left(1+{b_0V_0\over
V_1}\right)+\left({b_0V_0\over V_1}\right)^2\sqrt{1+2{V_1\over
b_0V_0}}={9\over 2} \Rightarrow d^2V_0=12 \label{81}\end{equation}

Using (\ref{80}), (\ref{79}) becomes
\begin{equation}
{3\over 8}V_0d^2 ={9\over 2}+{3(b_0^2V_0-12b_1V_0-27V_2)\over
5b_0V_0L}-{3b_1(-b_0^2V_0+12b_1V_0+27V_2)\over 5b_0^2V_0L^2}
\label{82}\end{equation}
$$-{
(64b_0^4-756b_0^2b_1-1944b_1^2-5400b_0b_2)V_0^2+2349b_0^2V_0V_2-20898b_1
V_0V_2-37179V_2^2+18225 b_0V_0V_3\over 750b_0^2V_0^2L^2}
$$
$$
+{\cal O}\left({1\over L^3}\right)
$$
We may calculate from (\ref{82})
\begin{equation}
{ d'\over d}=-{b_0^2V_0-12b_1V_0-27V_2\over
15V_0L^2}-{2b_1(b_0^2V_0-12b_1V_0-27V_2)\log L\over 15b_0V_0L^3}
\label{83}\end{equation}
$$
+{94b_0^4-1251b_0^2b_1-324b_1^2-5400b_0b_2\over 3375 b_0L^3}+{9(3b_0^2-31b_1)V_2\over 125 b_0V_0L^3}
-{567V_2^2\over 125 b_0 v_0^2L^3}+{27V_3\over 5V_0L^3}+\cdots
$$
Compatibility with the first of (\ref{70}) implies
\begin{equation}
{V_2\over V_0}={23b_0^2-36b_1\over 3^4}\sp {V_3\over V_0}=-2{324
b_2+124 b_0^3+189 b_0b_1\over 3^7} \label{84}\end{equation}
$$
{V_4\over V_0}={3176 b_0^4+7236b_0^2b_1-243 b_1^2-810 b_0b_2-4374 b_3\over 3^9}
$$
Using this we finally obtain for the metric
\begin{equation}
{V_0\over 12}d^2\equiv {d^2\over \ell^2}=1-{8b_0\over
3^2L}+{4(26b_0^2+9b_1)\over 3^4L^2}-{8b_1\log L\over 3^2L^2}-{8\over
3^2}{b_1^2\over b_0}{\log^2 L\over L^3} \label{85}\end{equation}
$$+{16b_1(13b_0^2+9b_1)\over 3^4b_0}{\log L\over L^3}
-{8\over 3^7 b_0}(698 b_0^4+594b_0^2b_1-243 b_1^2-324 b_0b_2){1\over L^3}
+{\cal O}\left({1\over L^4}\right)
$$
Therefore to next to leading order the Poicar\'e metric near the boundary is
\begin{equation}
ds^2=\left[1+{8\over 3^2\log {r\Lambda}}+{4\left(26+9{b_1\over
        b_0^2}-18{b_1\over b_0^2}\log(b_0\log {1\over r\Lambda})\right)\over 3^4\log^2 r\Lambda}+
{\cal O}\left ({\log^2\log {r\Lambda}\over \log^3 {r\Lambda}}\right)\right]
{\ell^2\over r^2}~(dr^2+d\vec x^2)
\end{equation}
We observe that the first non-trivial correction  is independent of $b_0$.
This is related to  the fact that as we have no unambiguous identification of the normalization of the gauge coupling,
the $\beta$-function coefficients we are using can be changed as
\begin{equation}
b_n\to a^{n+1}b_n
\label{scale}
\end{equation}
{}From (\ref{85}) we observe that $d$ is invariant up to the shift of the
logarithms, $\log L=\log (b_0\log{1\over r\Lambda})$
since it depends only on $b_1/b_0^2$ and $b_2/b_0^3$.
It is a  non-trivial statement that a change of scale in the coupling as
 in (\ref{scale}), in (\ref{85}) can be absorbed into a change of the scale $\Lambda$
as well as the non-universal $\beta$-function coefficients $b_{n>1}$.

The potential in the UV regime  can be therefore expanded in terms of the
 overall AdS scale as well as the $\beta$-function coefficients to obtain
\begin{equation}
V={12\over \ell^2}\left[1+{8\over 9}(b_0\l)+{23-36{b_1\over b_0^2}\over 3^4}(b_0\l)^2
-2{324{b_2\over b_0^3}+124+189{b_1\over b_0^2}\over 3^7}(b_0\l)^3+{\cal O}(\l^4)\right]
\end{equation}

\subsection{Scalar curvature invariants in the UV regime}

We may now evaluate the basic scalar invariants of the background in the weak coupling (UV) regime:

\begin{equation}
V=V_0\left[1+{8\over 9}{b_0\over L}+{23b_0^2\over 81L^2}+{8b_1\log
L\over 9L^2} +{\cal O}\left({1\over L^3}\right)\right]
\label{86}\end{equation}
\begin{equation}
(\partial \phi)^2=e^{-2A}{\dot\a^2\over \a^2}={r^2\over
d^2}{\dot\a^2\over \a^2}={1\over d^2}{ \a'^2\over \a^2}
=V_0\left[{b_0\over 12 L}+{b_0\log L\over 12 L^2}+{8b_0^2-9b_1\over
108L^2}+{\cal O}\left({1\over L^3}\right)\right]
\label{87}\end{equation}
\begin{equation}
R={4\over 3}(\partial \phi)^2-{5\over 3}V \label{88}\end{equation}
It is obvious that all invariants are regular in the UV, ( $L\to \infty$).
 All higher curvature  invariants are also regular as they are polynomial functions of
\begin{equation}
e^{-2A}\dot A^2={1\over 9}(\partial \phi)^2+{1\over 12}V\sp
e^{-2A}\ddot A=-{1\over 3}(\partial \phi)^2+{1\over 12}V
\label{89}\end{equation}

\subsection{The two-loop $\beta$-function coefficients of pure gauge theory}

As a final point we quote  the scheme-independent $\beta$-function
coefficients for QCD, $b_0,b_1$.

With  $N_f$ (non-chiral) flavors in the fundamental, the $\beta$-function reads
\begin{equation}
\beta(g)=-{g^3\over (4\pi)^2}\left\{{11\over 3}N_c-{2\over
3}N_f\right\} -{g^5\over (4\pi)^4}\left\{{34\over 3}N_c^2-{N_f\over
N_c}\left[{13\over 3}N_c^2-1\right]\right\}+\cdots \end{equation}

For the the 't Hooft
coupling, and with ${N_f\over N_c}\to x$ we obtain
\begin{equation}
\l\equiv {g^2 N_c}\sp \dot \l=-{2\over 3}\left[{(11-2x)\over (4\pi)^2}\l^2+{(34-13x)\over (4\pi)^4}\l^3+\cdots\right]
\end{equation}
from where we obtain
\be
b_0={2\over 3}{(11-2x)\over (4\pi)^2}\sp {b_1\over b_0^2}=-{3\over 2}{(34-13x)\over (11-2x)^2}
\ee
In this paper, $x=0$ and therefore ${b_1\over b_0^2}=-{3\cdot 34\over 2\cdot 121}\simeq 0.42$.

\section{General potentials at string tree level\label{dilpot}}

Extra terms in the dilaton potential can arise from higher $\alpha'$
corrections proportional to the field strength of the RR four-form.
We parametrize these corrections in the string frame as
\begin{equation}
\label{31}
S_F=-{M^{3}\over 2\ls^2}\int
d^5x\sqrt{g}~e^{-2\phi}K\left(e^{2\phi}y\right),
\end{equation}
where
\begin{equation}
K(y)\equiv
-2\dc+\sum_{n=1}^{\infty} {a_n\over n}y^n\sp y={\ls^2(F_5)^2\over
5!} \sp a_1=1\label{c2}
\end{equation}
and we assumed the simplest
type of contraction for simplicity. The contribution for
general type of contractions yield qualitatively the same result
upon modifying the coefficients, $a_n$. Therefore we will assume that (\ref{31}) captures all such corrections.

Passing to the Einstein frame we obtain,
\begin{equation}\lab{corractE}
S^E_F=-{M^3\over 2\ls^2}\int d^5x\sqrt{g}~e^{{4\over 3}\phi}K
\left(e^{-{14\over 3}\phi}y\right)
\end{equation}
The equations of motion are
\begin{equation}
\nabla^{\m_1}\left[e^{-{10\over 3}\phi}K'\left(e^{-{14\over
3}\phi}y\right) F_{\m_1\m_2\cdots\m_5}\right]=0
\label{32}\end{equation} with solution
\begin{equation}
K'\left(e^{-{14\over 3}\phi}y\right)
F_{\m_1\m_2\cdots\m_5}=N_ce^{{10\over 3}\phi}E_{\m_1\m_2\cdots\m_5}
\label{33}\end{equation}
where $N_c$ will now be non-linearly related to the number of color branes.
By squaring we obtain the consistency
conditions
\begin{equation}
y \left[K'\left(e^{-{14\over 3}\phi}y\right)\right]^2
=-N_c^2e^{{20\over 3}\phi}\sp (K')^2{F^2_{\m\n}\over 4!} =-{N_c^2\over
\ls^2}e^{{20\over 3}\phi}~g_{\m\n} \label{con}\end{equation}
so that
\begin{equation} F_{\m\n}^2={F^2\over
5}g_{\m\n} \label{35}\end{equation}
We now compute the contribution of the five-form to the stress tensor
\begin{equation} T_{\m\n}^F={1\over \sqrt{g}}{\delta S^E_F\over \delta
g^{\m\n}}=-{M^3\over 2} \left[e^{-{10\over 3}\phi}K'{F^2_{\m\n}\over
4!}-{g_{\m\n}\over 2\ell_s^2}e^{{4\over 3}\phi}K\right] ={M^3e^{{4\over
3}\phi}\over 2\ell_s^2}\left[-e^{-{14\over 3}\phi}yK'+{1\over
2}K\right]~g_{\m\n} \label{36}\end{equation}
We may therefore
substitute the action of the five form $S^E_F$ with
\begin{equation} \hat S^E_F={M^3\over
\ls^2}\int d^5x \sqrt{g}~V_E(\phi)\label{37}
\end{equation}
\be
 V_E(\phi)=e^{{4\over 3}\phi}
\left[e^{-{14\over 3}\phi}X(\phi)K'\left(e^{-{14\over
3}\phi}X(\phi)\right) -{1\over 2}K\left(e^{-{14\over
3}\phi}X(\phi)\right)\right]\label{34}
\ee
In the formulae above
$y=X(\phi)$ is a solution of (\ref{con}).

We redefine
\begin{equation}
\zeta=e^{-{14\over 3}\phi} y \label{2}
\end{equation}
Then the
relevant equations become
\begin{equation} \zeta [K'(\zeta)]^2=-N_c^2
e^{2\phi}\sp V_E=e^{{4\over 3}\phi}\left[\zeta K'(\zeta)-{1\over
2}K(\zeta)\right] \label{1}
\end{equation}

We return to the $\s$-model frame in order to estimate the large-$N_c$ dependence
\be
S_{\s}=M^3\int d^5x\sqrt{g}e^{-2\phi}\left[R+4(\partial \phi)^2+{1\over \ell_s^2}
V_{\s}(\phi)\right]\sp V_{\s}(\phi)=\left[\zeta K'(\zeta)-{1\over
2}K(\zeta)\right]
\label{1aa}\ee
 If we define the 't Hooft coupling as
\begin{equation}
\l\equiv N_c~e^{\phi} \label{39}\end{equation}
then from (\ref{1}), $\zeta$ is a function of $\l$.
The $\s$-model frame action (\ref{1aa}) becomes
\be
S_{\s}=N_c^2~M^3\int d^5x\sqrt{g}{1\over \l^2}\left[R+4{(\partial \l)^2\over \l^2}+{1\over \ell_s^2}V_{\s}(\l)\right]
\label{1aaa}\ee
We observe the ${\cal O}(N_c^2)$ dependence expected from the sphere action as well as the fact that
 there is a non-trivial potential for the 't Hooft coupling.

For small $\zeta$ we can  calculate the potential  perturbatively:
\begin{equation}
K(\zeta)=-2\dc+\zeta+{1\over 2}a_2\zeta^2+{\cal O}(\zeta^3)\sp \zeta
K'(\zeta)=\zeta+a_2\zeta^2+{\cal O}(\zeta^3)
\label{38}\end{equation}
to obtain
\begin{equation}\lab{ffr}
\zeta=-\l^2-2a_2\l^4+{\cal O}(\l^6)\sp V_{\s}=-{1\over 2}
\left(-2\dc+\l^2+{a_2\over 2}\l^4+{\cal O}(\l^6)\right)
\end{equation}
For a DBI-like example : $K(y)=2\sqrt{1+y}$ we obtain
\begin{equation}
\z=-{\l^2\over 1+\l^2}\sp V(\l)=-\sqrt{1+\l^2}
\label{40}\end{equation}

It is obvious from the above analysis that $V(\l)$ has an infinite
series of terms, that, after solving the equations of motion of the
four-form, are independent from $\alpha'$ except an overall
(universal) dependence in front. Therefore the higher-order in
$\alpha'$ corrections to the four-form are equivalent to a (leading)
order in $\alpha'$ potential. Surprisingly, this  potential has a
regular expansion at weak YM coupling.

This analysis obviously generalizes to the kinetic terms of the dilaton and graviton.
The  general action is of the form
\begin{equation}
S=\int d^5x\sqrt{g}e^{-2\phi}\left[V~R+4V_1(\pa \phi)^2+V_2\right]
\label{a41}\end{equation}
The functions  $V,V_{1,2}$ depend on $e^{2\phi}y$ with $y$ given in (\ref{c2}),
and summarize the higher $\alpha'$-corrections of the four-form.
Going through the same procedure as above, it can be shown that this
 is equivalent with integrating
out the four-form, and multiplying the kinetic terms with appropriate series in the exponential of the dilaton
as in the standard potential.
Moreover, higher derivative terms proportional to powers of the curvature as well as the dilaton derivatives will be generated.

\subsection{Corrections to the axion terms\label{axion-corr}}

The axion $a$, dual to the instanton density has a special position among the supergravity fields.
Its shift symmetry is protected, and its special normalization in the YM theory implies that its contributions
are suppressed by a power of $1/N_c^2$, due to the fact that the dual variable in YM is an angle, \cite{wit3}.
This is reflected in the fact that in string theory, $a$ is a RR field and therefore has suppressed dilaton dependence.
The leading term in the effective action is $\int \sqrt{g} ~(\partial a)^2$ in the string frame, and becomes
\be
\int \sqrt{g} e^{2\phi}(\partial a)^2={1\over N_c^2}\int \sqrt{g} \lambda^2~(\partial a)^2
\ee
in the Einstein frame where in the second equality we indicated the suppression of the $\theta$-induced vacuum energy.

We now consider the higher terms in the $\alpha'$ expansion that involve the 5-form field strength.
In analogy with the previous section we may write them in the string frame as\footnote{Higher terms
like $(\partial a)^4$ are suppressed by extra powers of $N_c$
and we do not need to consider them here.}
\begin{equation}
\label{31b}
S_{F,a}=-{M^{3}\over 2\ls^2}\int
d^5x\sqrt{g}~e^{-2\phi}\left[K_1\left(e^{2\phi}y\right)+K_2\left(e^{2\phi}y\right)e^{2\phi}(\partial a)^2\right],
\end{equation}
where
\begin{equation}
K_1(y)\equiv -2\delta c
\sum_{n=1}^{\infty} {a_n\over n}y^n\sp K_2(y)\equiv 1+
\sum_{n=1}^{\infty} {b_n\over n}y^n\sp y={\ls^2(F_5)^2\over 5!} \sp a_1=1\label{31c}
\end{equation}
Passing to the Einstein frame we obtain,
\begin{equation}
\lab{31d}
S^E_F=-{M^3\over 2\ls^2}\int d^5x\sqrt{g}~e^{{4\over 3}\phi}\left[K_1
\left(e^{-{14\over 3}\phi}y\right)+K_2
\left(e^{-{14\over 3}\phi}y\right)e^{{2\over 3}\phi}(\partial a)^2\right]
\end{equation}

The equations of motion for the 5-form are
\begin{equation}
\nabla^{\m_1}\left[e^{-{10\over 3}\phi}K_1'\left(e^{-{14\over
3}\phi}y\right)+e^{-{8\over 3}\phi}K_2'\left(e^{-{14\over
3}\phi}y\right)(\partial a)^2\right] F_{\m_1\m_2\cdots\m_5}=0
\label{31e}\end{equation} with solution
\begin{equation}
\left[K_1'\left(e^{-{14\over
3}\phi}y\right)+e^{{2\over 3}\phi}K_2'\left(e^{-{14\over
3}\phi}y\right)(\partial a)^2\right]
F_{\m_1\m_2\cdots\m_5}=N_ce^{{10\over 3}\phi}E_{\m_1\m_2\cdots\m_5}
\label{31f}\end{equation}

By squaring we obtain the consistency
conditions
\begin{equation}
y \left[K_1'\left(e^{-{14\over
3}\phi}y\right)+e^{{2\over 3}\phi}K_2'\left(e^{-{14\over
3}\phi}y\right)(\partial a)^2\right]^2
=-N_c^2e^{{20\over 3}\phi}\lab{31k}\end{equation}
and
\be
 \left[K_1'+e^{{2\over 3}\phi}K_2'(\partial a)^2\right]^2
{F^2_{\m\n}\over 4!} =-{N_c^2\over
\ls^2}e^{{20\over 3}\phi}~g_{\m\n} \label{31g}
\ee
so that
\begin{equation} F_{\m\n}^2={F^2\over
5}g_{\m\n} \label{31h}\end{equation}
We now compute the contribution of the five-form to the stress tensor
\begin{equation} -{2\over M^3}T_{\m\n}^F=-{2\over M^3}{1\over \sqrt{g}}{\delta S^E_F\over \delta
g^{\m\n}}=
 e^{-{10\over 3}\phi}\left[K_1'+e^{{2\over 3}\phi}K_2'(\partial a)^2\right]
{F^2_{\m\n}\over
4!}+{e^{2\phi}\over \ell_s^2}K_2\partial_{\mu}a\partial_{\nu}a-
\ee
$$
-{g_{\m\n}\over 2\ell_s^2}e^{{4\over 3}\phi}\left[K_1+e^{{2\over 3}\phi}K_2(\partial a)^2\right]
$$
\be
={e^{{4\over
3}\phi}\over \ell_s^2}\left[e^{-{14\over 3}\phi}y\left[K_1'+e^{{2\over 3}\phi}K_2'(\partial a)^2\right]-{1\over
2}\left[K_1+e^{{2\over 3}\phi}K_2(\partial a)^2\right]\right]~g_{\m\n}+{e^{2\phi}\over \ell_s^2}
K_2\partial_{\mu}a\partial_{\nu}a \label{31i}\end{equation}
The axion equation also reads
\be
\nabla^{\m}\left[e^{2\phi}K_2\nabla_{\m}a\right]=0
\ee
We redefine
\begin{equation}
\zeta=e^{-{14\over 3}\phi} y\sp \eta=e^{{2\over 3}\phi}(\partial a)^2 \label{31j}
\end{equation}
and (\ref{31k})
becomes
\be
\zeta [K_1'(\zeta)+K_2'(\zeta)\eta]^2=-N_c^2 e^{2\phi}
\lab{31m}\ee

The dual action which gives the same equations of motion is
\begin{equation} \tilde S^E={M^3\over
\ls^2}\int d^5x \sqrt{g}~e^{{4\over 3}\phi}\left[
\zeta(K_1'(\zeta)+\eta K_2'(\zeta))-{1\over 2}(K_1(\zeta)+\eta K_2(\zeta))\right]\label{31l}
\end{equation}
In the action above, $\zeta(\l,\eta)$ is a solution of (\ref{31m}).
In the string frame the dual action becomes
\begin{equation}
\tilde S_{\s}={M^3\over
\ls^2}\int d^5x \sqrt{g}~e^{-2\phi}\left[
\zeta(K_1'(\zeta)+\tilde\eta K_2'(\zeta))-{1\over 2}(K_1(\zeta)+\tilde\eta K_2(\zeta))\right]\label{31n}
\end{equation}
with $\tilde\eta=e^{2\phi}(\partial a)^2$.
We must  separate the kinetic term of  the axion from the higher derivative terms that appear because $\zeta$ depends non-trivially on $\eta$.
This turns out to be in the Einstein frame
\begin{equation} \tilde S^E_{axion-linear}=-{M^3\over
2\ls^2}\int d^5x \sqrt{g}~e^{2\phi} K_2(\zeta^*)(\partial a)^2\label{31o}
\end{equation}
where now $\zeta^*$ is a solution $\zeta K_1'(\zeta)=-N_c^2 e^{2\phi}$.

We finish this section by giving the most general dual action involving the five form
\be
S=\int d^5x\sqrt{g}~e^{-2\phi}~Z(e^{2\phi}y,z_i)
\label{31p}\ee
where $Z$ is an arbitrary function and $z_i$ are scalar invariants of other fields, $R, R^2, R_{\m\n}R^{\m\n}$, $(\partial \phi)^2$, $e^{2\phi}
(\partial a)^2)$ etc.

The dual action is given by the Legendre transform
\be
\tilde S=\int d^5x\sqrt{g}~\left[\zeta \partial_{\zeta}Z(\zeta,z_i)-{1\over 2} Z(\zeta,z_i)\right]
\label{31q}\ee
where $\zeta$ satisfies
\be
\zeta ~(\partial_{\zeta}Z(\zeta,z_i))^2=-\lambda^2
\label{31r}\ee

\subsection{Corrections to the gauge coupling constant identification\label{dilcoup}}

Consider now a probe D$_3$ brane and the coupling of the kinetic gauge field terms
\begin{equation}
S_{D_3}={T_3\over \ls^4}\int d^4x\sqrt{\hat
g}e^{-\phi}Z(e^{2\phi}y)Tr[F^2],\,\,
Z(y)=1+\sum_{n=1}^{\infty}{c_n\over n}y^n,\,\, y={\ls^2(F_5)^2\over 5!}
\label{41}\end{equation} where $T_3$ is dimensionless and $Z(y)$
summarizes higher-order couplings of the five form on the D-brane
(arising from disk diagrams). Going to the Einstein frame and
substituting from \ref{2},\ref{1} we obtain for the gauge coupling
constant
\begin{equation}
g_{YM}^2 ={e^{\phi}\over Z(\zeta)} \label{42}
\end{equation}
Therefore, these corrections, although without derivatives,  are due to the higher $\alpha'$ corrections on the branes.
A similar argument indicates that the tension of possible flavor branes
obtains similar  types of corrections.

\subsection{Other higher derivative corrections at the tree level\label{hd}}

After dualizing the five form, there remain true higher-derivative corrections associated to curvatures
and derivatives of the dilaton.
We will indicate here how such corrections can be instrumental in effectively
generating a constant term in the dilaton potential,
a fact that we have assumed in this paper in order to simplify our problem.

To give the idea, we will focus on the tree level, string frame effective action without
dilatonic kinetic terms and considered only as a function of the scalar
curvature (for simplicity)
\be
S=\int d^5x \sqrt{g}e^{-2\phi}~f(\ell_s^2R,e^{2\phi}y)
\ee
where we are using the same definition of $y$ as in (\ref{c2}) and $f$ is a
function with a regular series expansion around $R=y=0$ .
The equations of motion are
\be
\ell_s^2 f_R~R_{\m\n}-{1\over 2}g_{\m\n}f+e^{2\phi}\left[g_{\m\n}\square-
\nabla_{\m}\nabla_{\n}\right](e^{-2\phi}f)=0
\sp f_R\equiv {\delta f\over \delta R}
\ee
\be
\nabla^{\m}\left[f_y~F_{\m\n_1\cdots\n_4}\right]=0\sp  f_y\equiv {\delta f\over \delta y}
\ee

These equations admit an  $AdS_5$ solution,
\be
R_{\m\n}=-{4\over \ell^2_{AdS}}g_{\m\n}
\ee
 with $\phi=$constant and
\be
f_y~F_{\n_1\cdots\n_5}={N_{c}\over \ell_s}E_{\n_1\cdots\n_5}~~~\to~~~y~f_y^2\left[x,e^{2\phi}y\right]=-N_c^2
\ee
where we defined
\be
x=-20{\ell_s^2\over \ell^2}
\ee
The metric equation then becomes
\be
2xf_R\left[x,e^{2\phi}y\right]+5f\left[x,e^{2\phi}y\right]=0
\ee
and typically the two algebraic equations are expected to have a solution.
A simple example involves
\be
f=e^{2\phi}y+\ell_s^2R+\mu \ell_s^4R^2
\ee
which generates an $AdS_5$ space with
\be
{\ell^2\over \ell_s^2}={9\mu\over 10(7\pm\sqrt{49+180\mu\lambda^2})}\sp \lambda\equiv e^{\phi}N_c
\ee
In particular, a solution exists for $\lambda=0$ and it is
\be
{\ell^2\over \ell_s^2}={9\mu\over 140}
\ee

It is not difficult to verify that allowing now the dilaton to run logarithmically
 in the UV, so that $\lambda\to 0$ as $r\to 0$
is a small (subleading) perturbation in the $\alpha'$ expansion, generating a
 solution that logarithmically
asymptotes to the AdS solution above. In particular, the kinetic terms of the dilaton
that have been neglected will be suppressed by large inverse logs, and the same
 applies to the power series in the t' Hooft coupling $N_ce^{\phi}$.
  We will not pursue this avenue further here, leaving it for a future investigation.

\section{Perturbative analysis  near an
extremal (AdS) point of the dilaton potential\label{pertpot}}

In this appendix we will analyze here the stability properties of the perturbative dilaton potential.

We will use the domain-wall coordinate, (\ref{dw}).
The field equations are given by (\ref{eins1}), (\ref{eins2}) and (\ref{eom2}).
We parametrize the potential around an AdS extremum as
\begin{equation}
V={12\over \ell^2}-{16\xi\over 3\ell^2}\Phi^2+{\cal O}(\Phi^3)
\label{param}
\end{equation}
where $\Phi<<1$.
Since $V'=0$ at the critical point, there is a $AdS_5$ solution
with
\begin{equation}
A={u\over \ell}\sp \Phi=0
\end{equation}
Perturbing around the fixed point solution $A=A_*+\delta A$,
$\Phi=\delta \Phi$ we obtain to linear order
\begin{equation}
{18\over \ell}\delta A'=\delta\Phi'^2-{4\over \ell^2}\Phi^2={\cal O}(\delta\Phi^2)
\sp \delta\Phi''-{4\over \ell}\delta\Phi'-{4\xi\over
\ell^2}\delta\Phi=0
\end{equation}
We observe that to linear order $\delta A$ is a constant, which amounts to a
 renormalization of the AdS length scale $\ell$.
We can therefore ignore it.
The general solution of the second equation is
\begin{equation}
\delta\Phi=C_+e^{{(2+2\sqrt{1+\xi})u\over \ell}}+
C_-e^{{(2-2\sqrt{1+\xi})u\over \ell}}
\end{equation}
Changing variable to $r=Le^{u/\ell}$ we obtain to linear order in the perturbation
\begin{equation}
ds^2={\ell^2\over r^2}(dr^2+d\vec x^2)+\cdots\sp \delta\Phi=C_+\left({r\over
\ell}\right)^{(2+2\sqrt{1+\xi})}+
C_-\left({r\over
\ell}\right)^{(2-2\sqrt{1+\xi})}
\end{equation}

\begin{figure}
 \begin{center}
 \leavevmode \epsfxsize=7cm \epsffile{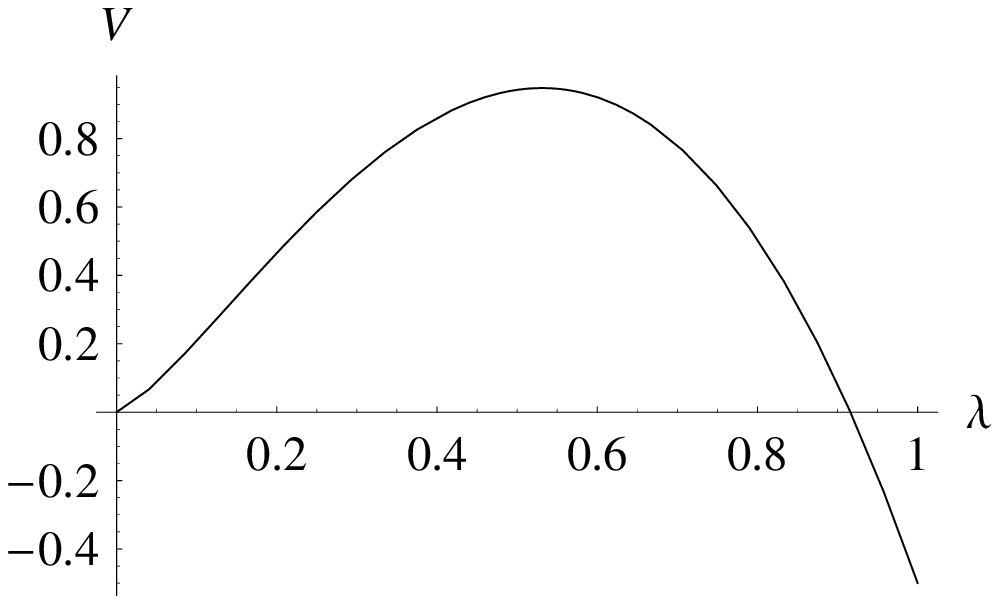}
  \leavevmode \epsfxsize=7cm \epsffile{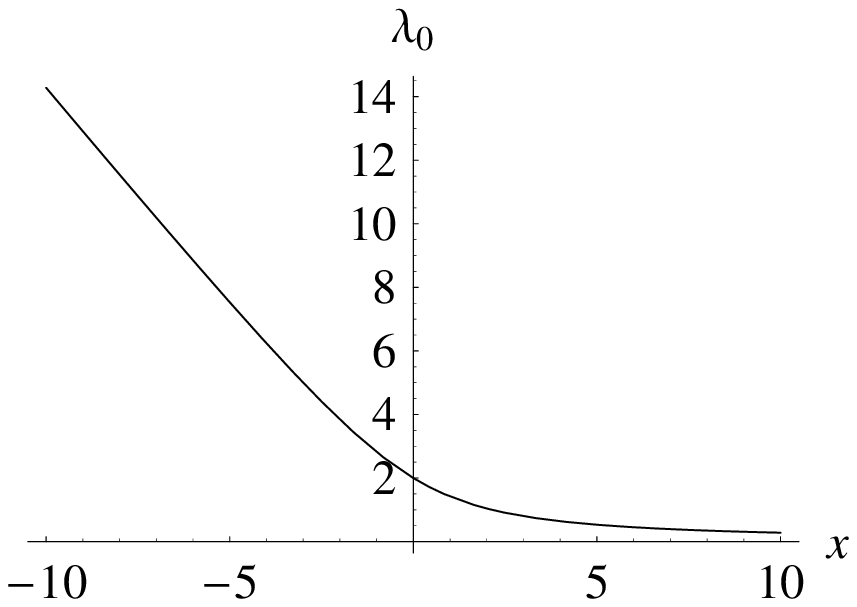}
 \end{center}
 \caption[]{Left:The dilaton potential  plotted as a function of the t 'Hooft coupling for $x=5$.
 Right: The value of the t 'Hooft coupling at the extremum as a function of $x$. }
 \label{f0}\end{figure}

We may now analyze the potential (we assume we are in 5 dimensions and therefore $\dc=5$)
\begin{equation}\lab{pot}
V(\Phi)={ \lambda^{\frac43} \over \ls^2}\left[5-\half \lambda^2
-x~\lambda\right]
   \end{equation}
where
\be\lab{defxf}
x\equiv{N_f\over N_c}.
\ee
   The potential is plotted in figure \ref{f0}

It has a single extremum at $\lambda=\lambda_0$ with:
\be
\lambda_0\equiv N_c e^{\phi_0}={-7x+ \sqrt{49 x^2+400}\over 10}.
\ee
{}From this we obtain the value of the potential at the extremum that gives,
\be
{\ell^2_s\over \ell^2}= \frac{\lambda_0^{\frac43}}{400}\left[{100 +7x^2-x\sqrt{49 x^2+400}\over 400}\right]
\ee
as well as the second derivative parametrized as in (\ref{param})
\be
\xi={5\over 4}\left[{400 +49x^2-7x\sqrt{49 x^2+400}\over {100+7x^2-x\sqrt{49 x^2+400}}}\right]
\ee

\begin{figure}
 \begin{center}
 \leavevmode \epsfxsize=7cm \epsffile{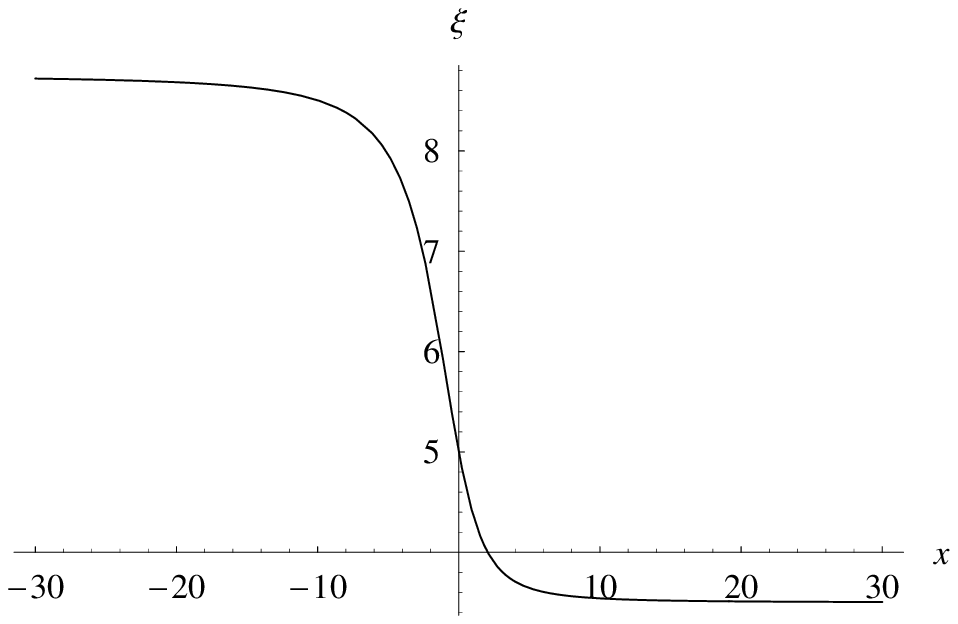}
  \leavevmode \epsfxsize=7cm \epsffile{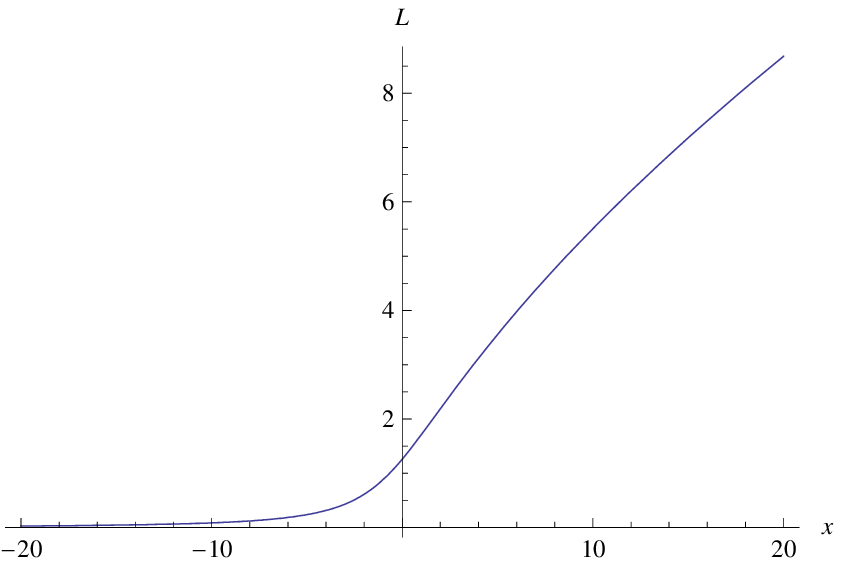}
 \end{center}
 \caption[]{Left:$\xi$   plotted as a function of $x$.
 Right: The value of the AdS radius $L$ in units of $\ell_s $ as a function of $x$. }
 \label{f00}\end{figure}

As $x$ varies between $0\leq x\leq \infty$, the parameter $\xi$ that controls
the anomalous dimension of the YM coupling constant
varies as $5\geq \xi \geq {7\over 2}$ as can be seen in figure \ref{f00}.

This implies that the associated dimension is $\Delta=2+2\sqrt{1+\xi}$ and
satisfies $2+3\sqrt{2}<\Delta<2+2\sqrt{6}$ or equivalently
$6.24<\Delta<6.90$.
It therefore corresponds to an irrelevant operator. This is not what we expect
 for QCD at weak coupling.

The AdS radius $\ell$ (in unite of $l_s$) becomes small at $x<0$, whereas it increases
as $x$ becomes large as can be seen in figure \ref{f00}.
Therefore it becomes arbitrarily larger than the string scale for large $x$ and therefore
the background will be weakly curved.
Therefore, although the t 'Hooft coupling at the AdS extremum
is small as $x$ becomes large, the geometry is weakly curved.
This is against the AdS$_5\times$S$^5$ intuition.

It should also be noted that the coefficients $N_c$ and $N_f$ appearing in the potential
 are multiplicatively related to the number of
colors and flavors respectively. They equal them if the relations stemming from
${\cal N}=4$ branes still hold, but this is not in general guaranteed.

\section{Analysis of the dilaton potential with single exponential}
\lab{exponential}

In the string inspired dilaton potential, the weak coupling asymptotics of the
flow is governed by a single exponential in the potential. Also in
the confining examples that we study in this paper, the leading term
of the potential in the IR is an exponential. Therefore it is
desirable to investigate the solutions of the system given by a
potential of the following form (after a convenient shift in the
dilaton)
\def\f{\phi}
\begin{equation}\label{sxpot}
    V(\f) = \frac43 \e\, e^{\alpha\f}, \quad \e = \pm 1.
\end{equation}

We want to find all the solutions of the system of equations,
(\ref{eins1}) and (\ref{eom2}) with $V$ given by (\ref{sxpot}). As a
starting point, we classify the solutions by the behavior of the
phase space variable $X$ which, in this case, obeys the following simple
equation obtained from (\ref{pstgen1}):
\begin{equation}\label{sxps}
  \frac{dX}{d\f}=\frac43(X+\frac1a){(X^2-1)\over 6X},
\end{equation}
where we defined
\be\lab{deffa}
a=8/3\alpha.
\ee
The fixed points of (\ref{sxps}) are
given by $X= \pm 1$ and $X = -1/a$.

Equation (\ref{sxps}) can be integrated to yield
\begin{equation}
e^{\phi-\phi_0}={(X-1)^{3a \over 8(1+a)}(X+1)^{3a\over 8(1-a)}\over
\left(X+ \frac1a \right)^{3a\over 4(1-a^2)}}.
\label{47}\end{equation} For the special case of $a=\pm 1$
we obtain instead
\begin{equation}
\log{X-1\over X+1}-{2\over X\pm 1}={16\over 3}(\phi-\phi_0).
\label{48}\end{equation}

Now, the solutions of the system are given by the fixed point
solutions of (\ref{sxps}) and the solutions that flow between these
fixed points. We first list the fixed point solutions:

\subsection{The $X=1$ fixed point}

This amounts to solving
\begin{equation}
\phi'=3 A'\sp \phi''+{4\over 3}\phi'^2=0\sp e^{\alpha\phi}\to 0
\label{A7}\end{equation} with solution
\begin{equation}
e^{\phi}=C\left(u_0-u\right)^{3\over 4} \label{A8}\end{equation}
We take the range of $u$ as $u\in(-\infty,u_0)$.
If $\alpha>0$ this solution is valid in the immediate neighborhood of
$u=u_0$. If $\alpha<0$ it is valid in the neighborhood $u\to \infty$. We
also find
\begin{equation}
e^{A}=\tilde C(u_0-u)^{1\over 4} \label{A9}\end{equation} and for
the Poincar\'e coordinate
\begin{equation}
r-r_0={4\over 3\tilde C}\left(u_0-u\right)^{3\over 4}\sp
b(r)=\left({3\tilde C^4\over 4}\right)^{1\over 3}(r-r_0)^{1\over 3}
\sp e^{\phi}={3\over 4}C\tilde C (r-r_0) \label{A10}\end{equation}
This solution is valid near $r=r_0$.

\subsection{The $X=-1$ fixed point}

This amounts to solving
\begin{equation} \phi'=-3 A'\sp \phi''-{4\over 3}\phi'^2=0\sp
e^{\alpha\phi}\to 0 \label{A11}\end{equation} It is related to the $X=1$
solution by $\phi\to -\phi$.

\subsection{The $X=-\frac1a$ fixed point}

 It exists if

 (a) $\e=1$ and $|a|> 1$

 (b) $\e=-1$ and $|a|< 1$

 (c) $\e=\pm 1$   and  $|a|=1$. In this case it merges
 with the $X=\pm 1$ fixed point solutions.

The solution is
\begin{equation}
e^{\phi}=\left({C\over u_0-u}\right)^{-{3a\over 4}}\sp e^{A}=\tilde
C \left(u_0-u\right)^{{a^2\over 4}}\sp C={3a\over
  4}\sqrt{\e\left(a^2-1\right)}
\label{A14}\end{equation}

In the Poincar\'e coordinate,
\begin{equation}
r-r_0={4\over (4-a^2)\tilde
C}\left(u_0-u\right)^{4-a^2\over 4}\sp e^A{(r)}=\tilde
C\left({(4-a^2)\tilde C\over 4}\right)^{a^2\over 4-a^2}
(r-r_0)^{a^2\over 4-a^2} \label{A15}\end{equation}
\begin{equation}
e^{\phi}=C^{-{3a \over 4}}\left({(4-a^2)\tilde C\over 4}
\right)^{3a\over 4-a^2}(r-r_0)^{3a\over 4-a^2}
\label{A16}\end{equation}

\subsection{Flow solutions}

\begin{figure}
 \begin{center}
  \leavevmode \epsfxsize=14cm \epsffile{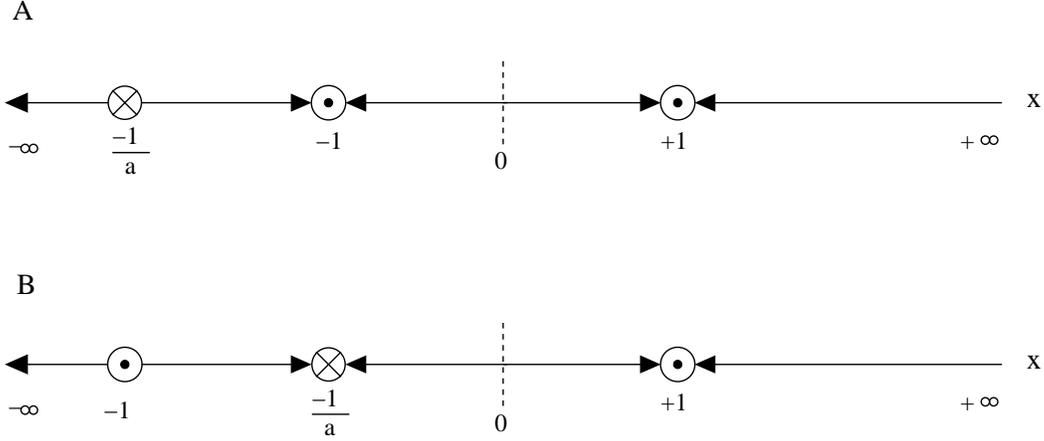}
 \end{center}
 \caption[]{The stability properties of the fixed points. Case A is for $0<a<1$
   and case B corresponds to $a>1$. The cases where $a<0$ can be obtained from
 the above figures by using the reflection symmetry in $X$. }
 \label{flow}\end{figure}

More general solutions are given by the flows between the fixed
points of $X$. The nature of these flows are determined by the
stability properties of the fixed points that we depict in
fig.(\ref{flow}) for the case $a<0$. A cross denotes an unstable
fixed point and a point denotes a stable one. The arrows between
these points depict the direction of the flow in an appropriate
radial variable that we define below. The boundaries of the phase
space are given by $X=\pm \infty$ and $X=0$. In particular the $X<0$
and the $X>0$ solutions are disconnected. The case $a>0$ can be
obtained from fig.(\ref{flow}) by utilizing the symmetry of
(\ref{sxps}) under $X\to -X$, $\f\to -\f$ and $a\to -a$.

The system can be solved analytically in the following radial
variable, $t$:
\begin{equation}\label{tvar}
    \frac{d}{du} = e^{\frac{\a}{2}\f} \frac{d}{dt}
\end{equation}

We describe the solution for the case $\e = +1$ which is the
interesting case for us. In this case the range of the phase space
variable is given by $-1<X<1$. This range is divided into two flow
parts by the presence of the fixed point at $X = -1/a$. Let us take
$a>1$ ($a<1$ is given by a simple replacement of the hyperbolic
functions below with the triangular ones).

\subsubsection{The solution: $-1<X<-1/a$}

Let us define,
\begin{equation}\label{tbar}
\tb = \frac{2\sqrt{a^2-1}}{3a}(t-t_0)
\end{equation}
 The solution is given by,
\begin{equation}\label{flow1f}
    \f = \f_0 + \frac{3a}{4}\le(\frac{1}{a+1}\log\cosh(\tb)
    -\frac{1}{a-1}\log\sinh(\tb)\ri),
\end{equation}
and
\begin{equation}\label{flow1a}
    A = A_0 +\frac{a}{4}\le(\frac{1}{a+1}\log\cosh(\tb)
    +\frac{1}{a-1}\log\sinh(\tb)\ri).
\end{equation}
The phase space variable as a function of $\tb$ is given by,
\begin{equation}\label{flow1x}
    X(\tb) =
    \frac{\frac{a-1}{a+1}\tanh^2(\tb)-1}{\frac{a-1}{a+1}\tanh^2(\tb)+1}.
\end{equation}

In these equations, we choose $\tb>0$ that corresponds to the range
$t_0<t<\infty$. This solution corresponds to the flow from the fixed
point $X=-1$ at $t=t_0$ to the fixed point $X = -1/a$ at $t=\infty$.

The asymptotics of this solution are as follows. As $t\to t_0$,
($X\to -1$),
\begin{eqnarray}
  \l &\to& (t-t_0)^{\frac{3a}{4(a-1)}} \\
  ds^2 &\to& (t-t_0)^{\frac{a}{2(a-1)}}dx^2+(t-t_0)^{\frac{2}{a-1}}dt^2
\end{eqnarray}
Thus the space shrinks to a point as one approaches to the $X=-1$
fixed point. As $t\to\infty$, ($X\to -1/a$),
\begin{eqnarray}
  \l &\to& e^{-\frac{t}{\sqrt{a^2-1}}} \label{as1}\\
  ds^2 &\to& \frac{1}{r^2}\le(dx^2 +
  r^{-\frac{8}{a^2}}dr^2\ri)\label{as2}
\end{eqnarray}
where in the last equation we changed the variable as $r =
\exp(-at/\sqrt{a^2-1})$, hence $r\to 0$ in the limit. We observe
from the asymptotics of the metric, that the space becomes AdS only
when $a$ is taken to $\infty$. This nicely parallels the fact that
as $a\to\infty$ the fixed point approaches to $X\to 0$ and, as our
primary example in this paper suggests, one has an asymptotically AdS
space in the $X\to 0$ limit.

\subsubsection{The solution: $-1/a<X<+1$}

 The solution is given by,
\begin{equation}\label{flow2f}
    \f = \f_0 + \frac{3a}{4}\le(-\frac{1}{a-1}\log\cosh(\tb)
    +\frac{1}{a+1}\log\sinh(\tb)\ri),
\end{equation}
and
\begin{equation}\label{flow2a}
    A = A_0 +\frac{a}{4}\le(\frac{1}{a-1}\log\cosh(\tb)
    +\frac{1}{a+1}\log\sinh(\tb)\ri).
\end{equation}
The phase space variable as a function of $\tb$ is given by,
\begin{equation}\label{flow2x}
    X(\tb) =
    \frac{-\frac{a+1}{a-1}\tanh^2(\tb)+1}{\frac{a+1}{a-1}\tanh^2(\tb)+1}.
\end{equation}
This solution is valid in the whole range $-1/a<X<+1$ (for $a>1$)
and describes the two flow solutions: one from $X=0$ to $X = -1/a$
and from $X=-1/a$ and one from $X=0$ to $X=1$.

The interesting asymptotics are $t\to\infty$, ($X\to-1/a$) and $t\to
t_1$ where,
\begin{equation}\label{t1}
    t_1= t_0+\frac{3a}{2\sqrt{a^2-1}}\tanh^{-1}\sqrt{\frac{a-1}{a+1}}.
\end{equation}

In the first case, as $X\to-1/a$, one obtains the same asymptotics
as in (\ref{as1}) and (\ref{as2}). In the second case, as $X\to 0$,
one finds,
\begin{eqnarray}
  \l &\to& const. \\
  ds^2 &\to& dx^2+ dt^2.
\end{eqnarray}

\section{$\beta$-function with an exponential tail\label{bapp} }

In this appendix we provide another example of an exact $\beta$-function and the associated holographic geometry.
We choose the $\beta$-function here as:
\be
\b=-b_0 \l^2 e^{-c_0\l}
\ee
In perturbation theory this describes an asymptotically free theory. In the IR it reaches a fixed point but
in way which is different from the previous example. In particular the fixed point is at infinite coupling.

The potential that follows from this $\beta$-function is obtained from
(\ref{potx}) as,
\begin{equation}\lab{V3s}
V =
V_0(1-\frac{b_0^2}{9}\l^2e^{-2c_0\l})e^{-\frac{8b_0}{9c_0}\le(e^{-c_0\l}-1\ri)
},
\end{equation}

\begin{figure}
 \begin{center}
 \leavevmode \epsfxsize=12cm \epsffile{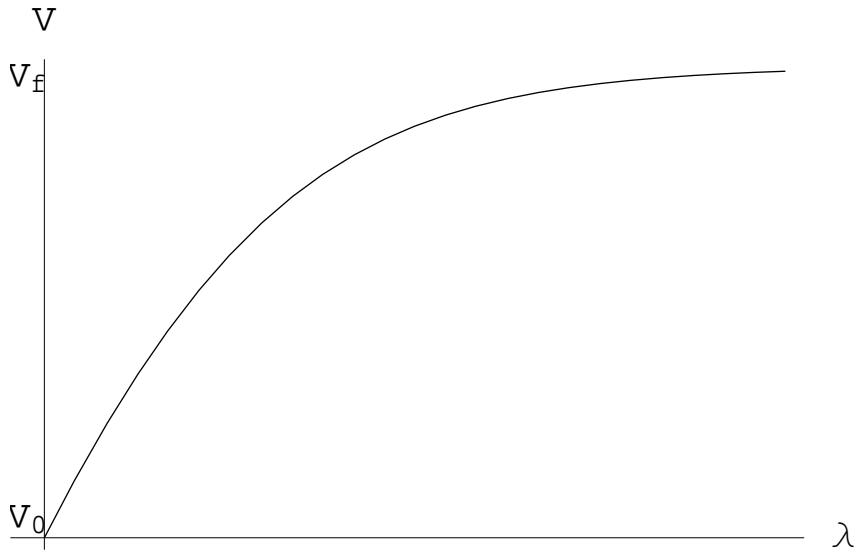}
 \end{center}
 \caption[]{Scalar potential as a function of $\l$ for the exponential type running.
We set $V_0=b_0=c_0=1$ for illustrative purposes. }
 \label{fig5}\end{figure}

The plot of the potential is shown in fig. \ref{fig5}. The RG flow
is from the Gaussian fixed point in the UV to an IR fixed point at
$\l=\infty$ in the IR. The corresponding UV geometry is an
asymptotically AdS space with radius given by (\ref{rads}) as
before. The corresponding far IR geometry is also AdS as expected from the
fact that $V$ in (\ref{V3s}) approaches to a constant as $\l\to\infty$:
\be\lab{expVf}
V\,\,\to\,\, V_f = V_0 e^{\frac{8b_0}{9c_0}}.
\ee
The radius of the AdS in the IR is given by,
\begin{equation}\label{IRrads2}
    \ell_{IR} = 2\sqrt{\frac{3}{V_f}}.
\end{equation}

\begin{figure}
 \begin{center}
 \leavevmode \epsfxsize=12cm \epsffile{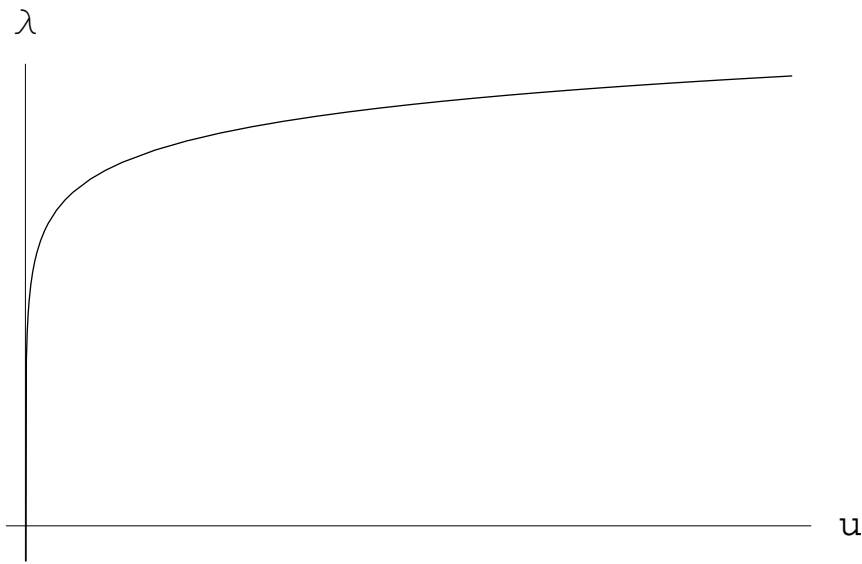}
 \end{center}
 \caption[]{Running of the coupling constant in the exponential case.
We have set $V_0=b_0=c_0=1$.}
 \label{fig6}\end{figure}

The running of the 't Hooft coupling is determined by the following
differential equation that follows from (\ref{fprime}):
\begin{equation}\lab{l3s}
\l' =
\frac{\sqrt{3V_f}}{6}b_0\l^2\,e^{-c_0\l}\,e^{-\frac{4b_0}{9c_0}\le(e^{-c_0\l}-1\ri)}.
\end{equation}
We display the numerical solution to this equation in figure
\ref{fig6}. From this figure it is clear that the coupling constant
diverges in the IR. One can calculate the
scalar invariants of the geometry.
The Ricci scalar is given by (\ref{E}) and $R = -3/2 E$. In the
Einstein frame, one obtains the IR limit value of $R$ as,
\begin{equation}\lab{r4}
  R = -\frac{15}{4}V_f,
\end{equation}
where $V_f$ is given by (\ref{expVf}).
Note that the function $X$
in (\ref{defx}) limits to zero both in the UV and the IR. This is in accord
with the fact that the scalar potential goes over to constant values both  in
the UV and the IR. The 5D geometry is
of the form of a Minkowski domain wall that interpolates between two
AdS geometries.

The details of the approach towards the IR AdS geometry can be
studied as in section \ref{bankszaks}. One determines the following
subleading behavior near the IR fixed point:
\begin{equation}\label{expIR}
    ds^2 =
    \frac{l_{IR}^2}{r^2}(1-\frac49\frac{1}{r\Lambda}\frac{1}{\log\log(r\Lambda)})^{-2}\le(dx^2+dr^2\ri),
\end{equation}
as $r\to\infty$.

\end{document}